\documentclass[structabstract]{aa}  
\usepackage{graphicx}
\usepackage{txfonts}
\usepackage{longtable,lscape}
\begin{document}
\def\teff{$T_{\rm eff}$}
\def\kms{{km~s}$^{-1}$}
\def\logg{$\log g$}
\def\turb{$\xi$}
\def\rad{$V_{\rm r}$}
\def\elem{$\varepsilon ({\rm El})$}
\def\vsini{$V\sin i$}
\title{Abundance analysis of prime B-type targets for~asteroseismology\thanks{Atmospheric parameters and abundance analysis results are available only in electronic form at the CDS via anonymous ftp.}}
   \subtitle{II. B6--B9.5 stars in the field of view of the {\it CoRoT}\thanks{{The {\it CoRoT}
   space mission was developed and is operated by the French space agency CNES,
   with participation of ESA's RSSD and Science Programmes, Austria, Belgium,
   Brazil, Germany, and Spain.}} satellite} \titlerunning{Abundance analysis of
   prime B-type targets for asteroseismology. II}

   \author{E. Niemczura
          \inst{1,2}
          T. Morel
          \inst{2,3}
          C. Aerts
          \inst{2,4}
          }

   \institute{Astronomical Institute, Wroc{\l}aw University, Kopernika 11, 51-622 Wroc{\l}aw, Poland\\
              \email{eniem@astro.uni.wroc.pl}
              \and
              Institute of Astronomy, Katholieke Universiteit Leuven, Celestijnenlaan 200D, 3001 Leuven, Belgium\\
              \email{conny@ster.kuleuven.be}
              \and
              Institut d'Astrophysique et de G\'eophysique, Universit\'e de Li\`ege, All\'ee du 6 ao\^ut, B\^at. B5c, 4000 Li\`ege, Belgium\\
              \email{morel@astro.ulg.ac.be}
              \and
              Department of Astrophysics, IMAPP, Radboud University Nijmegen, PO Box 9010, 6500 GL, Nijmegen, The Netherlands\\
             }

   \date{Received ...; accepted ...}


  \abstract
{The CoRoT satellite is collecting precise time-resolved photometry for tens of asteroseismology targets. To ensure a correct  interpretation of the CoRoT data, the atmospheric parameters, chemical compositions, and rotational velocities of the stars must be determined.}
{The main goal of the ground-based seismology support program for the CoRoT mission was to obtain photometric and spectroscopic data for stars in the fields monitored by the satellite. These ground-based observations were collected in the GAUDI archive. High-resolution spectra of more than 200 B-type stars are available in this database, and about 45\% of them is analysed here.}
{To derive the effective temperature of the stars, we used photometric indices. Surface gravities were obtained by comparing observed and theoretical Balmer line profiles. To determine the chemical abundances and rotational velocities, we used a spectrum synthesis method, which consisted of comparing the observed spectrum with theoretical ones based on the assumption of LTE.}
{Atmospheric parameters, chemical abundances, and rotational velocities were determined for 89 late-B stars. The dominant species in their spectra are iron-peak elements. The average Fe abundance is 7.24$\pm0.45$\,dex. The average rotational velocity is 126\,\kms, but there are 13 and 20 stars with low and moderate \vsini\, values, respectively. The analysis of this sample of 89 late B-type stars reveals many chemically peculiar (CP) stars. Some of them were previously known, but at least 9 new CP candidates, among which at least two HgMn stars, are identified in our study.
These CP stars as a group exhibit \vsini \ values lower than the stars with normal surface chemical composition.
}
   {}


   \maketitle
%


\section{Introduction}
\onltab{1}{}
\onltab{2}{}
In preparation for the {\it CoRoT} space mission (see {\it The CoRoT Book}, Fridlund et al. \cite{fridlund}), an extensive ground-based photometric and spectroscopic campaign of the stars located in the field of view of the satellite was performed. All of these data were collected in the {\it Ground-based Asteroseismology Uniform Database Interface} (GAUDI, Solano et al. \cite{solano}). This catalogue contains a vast number of spectra of B-type stars, among others.

Only a few investigations of B stars in the GAUDI archive have been completed. Neiner et al. (\cite{neiner}) discovered 17 new Be stars in the GAUDI database. Fr\'emat et al. (\cite{fremat}) determined the fundamental parameters (spectral type, temperature, gravity, \vsini) of the 64 Be stars in the seismology fields of {\it CoRoT} by performing a detailed modelling of the stellar spectra. Lefever et al. (\cite{lefever2006}) and Lefever (\cite{lefever2007}) estimated the fundamental stellar parameters, i.e., effective temperatures, surface gravities, luminosities, He and Si abundances, and projected rotational velocities of a subsample of B stars in GAUDI. They used line-profile fitting methods based on non-LTE line-formation predictions for unified atmospheres with winds. Morel \& Aerts (\cite{morelaerts2007}) presented an abundance study of two early B-type targets for the asteroseismology programme of the {\it CoRoT} mission. 

To provide a homogeneous high-precision determination of the fundamental parameters and abundances of the prime B-type targets of the {\it CoRoT} space mission, Morel et al.\ (\cite{morel2006}; Paper I) performed a pilot study of nine bright prototypical early-B $\beta\,$Cephei stars. This study was based on Kurucz (\cite{kurucz}) atmosphere models and non-LTE modelling techniques, as well as a time series analysis of high-quality spectroscopic data assembled for these stars primarily for spectroscopic mode identification. In a follow-up study, Morel et al.\ (\cite{morelhubrig2008}) increased the sample to 20 stars. These studies unexpectedly discovered a population of nitrogen-rich and boron-depleted, yet intrinsically slowly-rotating early-B stars (see also, e.g., Gies \& Lambert \cite{gieslambert}). A higher incidence of chemical peculiarities was also found in the stars for which a weak magnetic field was detected. Although this needs to be confirmed by analysing larger samples, this suggests that magnetic phenomena may be important in altering the photospheric abundances of early B dwarfs, even for weak surface field strengths.

As pointed out by Morel \& Aerts (\cite{morelaerts2007}), there is a large amount of spectroscopic data of B-type stars in the GAUDI archive. Most of these stars are late B-type objects. In this paper, we present a homogeneous study of 89 B6--B9.5 stars in the field of view of the {\it CoRoT} mission. The chemical abundances that we derive here provide useful additional information for the interpretation of the {\it CoRoT} data. 

The atmospheres of late-B main-sequence stars are relatively simple to model. There is no convection, stellar winds are very weak, and for the majority of these objects the microturbulence is very low. In these conditions, diffusion can easily occur. A significant number of late-B stars exhibit spectroscopic signatures of these microscopic transport processes (e.g., Hempel \& Holweger~\cite{hempel2003}; Folsom et al.~\cite{folsom}). The investigation of late-B star atmospheres will help us to understand these physical phenomena. It will provide valuable information about the behaviour of an element influenced by both gravitation and radiation, leading to a variety of abundance patterns even in nonmagnetic stars. Our goal is to investigate these effects for a large number of stars.

This paper is organised as follows. In Sect.\,2, we describe the spectroscopic observations used in the abundance analysis. In Sect.\,3, we present the method for determining the atmospheric parameters, abundance patterns and projected rotational velocities of the objects. The sources of possible errors in the chemical abundances are described in Sect.\,4. The measured \vsini\, values and elemental abundances are discussed in Sects.\,5 and 6, respectively. Section\,7 contains a detailed discussion of the chemical abundances of known and newly discovered chemically peculiar stars. Conclusions and future prospects are given in Sect.\,8. Reliability tests of our abundance analysis method are presented in Appendix\,A.

\section{Spectroscopic observations}
We analyse the high-resolution ($R \cong 40\,000-50\,000$) spectroscopic observations of B-type stars obtained with the ELODIE and FEROS spectrographs (see Table\,1). The ELODIE spectrograph (Baranne et al. \cite{baranne}) was attached to the 1.93\,m telescope at the Observatoire de Haute-Provence, France. The FEROS spectrograph (Kaufer et al. \cite{kaufer}) was installed on the 1.52\,m and 2.2\,m telescopes at La Silla, Chile. Both instruments are cross-dispersed, fibre-fed echelle spectrographs. Typical signal-to-noise ratios of the spectra range from 100 to 150 at 5500\,{\AA}. The wavelength intervals covered are 3900--6800\,{\AA} and 3800--9100\,{\AA} for ELODIE and FEROS, respectively. For the details of the data reduction, we refer to Solano et al.~(\cite{solano}). In our initial sample, we have 160 FEROS spectra and 62 ELODIE spectra of B-type stars. Only the spectroscopic observations of the B6--B9.5 stars available in the GAUDI archive are analysed in this paper. We also excluded Be stars, supergiants and spectroscopic binaries for which the lines of two stars were visible in the spectra. Basic information about the analysed stars, including spectral type, name of the spectrograph, observation date in UT, V magnitude, interstellar reddening,  indications of binarity, and abundance determinations in the literature are given in Table\,1.\footnote{Table\,1 and Table\,2 are only available in the electronic form.}

A comparison between the Balmer line profiles for stars with both FEROS and ELODIE spectra in GAUDI showed that the wings are far less pronounced in the latter (see Lefever 2007), whereas good agreement was found with ELODIE spectra taken from the archives \footnote{http://atlas.obs-hp.fr/elodie/}. Although the origin of the problem affecting the ELODIE spectra in GAUDI is unclear (but probably related to the data reduction), using these data would produce spuriously low gravity estimates. We therefore only used archival ELODIE spectra in our analysis. All the spectra were carefully normalised by us using standard IRAF procedures. This was found to be necessary to guarantee high-precision estimates of the fundamental parameters, especially the gravity. The extraction of Balmer line profiles from echelle spectra can be problematic, because in many cases the wings extend beyond the spectral coverage of a single order in the echellogram or the line is situated at the edge of the order. This is the situation for the FEROS and ELODIE spectrographs. In these cases, the proper setting of the line intensities depends not only on the continuum placement process but also on the data reduction (flat-fielding, order connections). For this reason, all Balmer lines suitable for determining the surface gravity were analysed (see Sect.\,3). We found that there is a high degree of consistency between the $\log g$ derived from different lines, typically $\Delta \log g < $\,0.2.

\section{Method of analysis}

   \begin{figure}
   \centering
   \includegraphics[width=8cm]{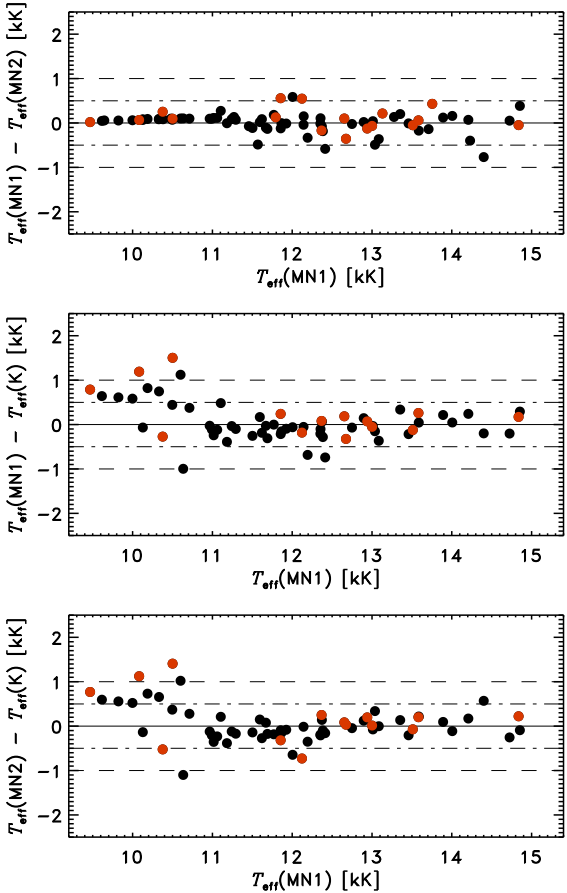}
      \caption{Comparison between the effective temperatures calculated from Str\"{o}mgren and Geneva photometry. The \teff(MN1) and \teff(MN2) symbols indicate \teff\, values determined by the two methods evaluated by Napiwotzki et al. (\cite{napiwotzki}), \teff\,(K) corresponds to effective temperatures obtained from Geneva photometry (K\"{u}nzli et al. \cite{kunzli}). Red dots denote confirmed or suspected CP stars.}
   \label{figure1}
   \end{figure}

   \begin{figure}
   \centering
   \includegraphics[width=9cm]{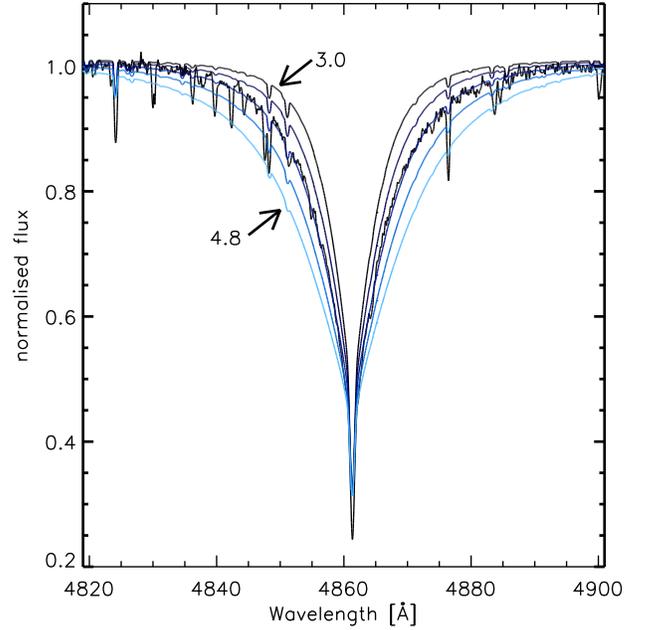}
      \caption{Surface gravity determination from Balmer lines. The observed H$\beta$ profile of \object{HD\,46886} is shown together with synthetic profiles obtained for different \logg\, values (from 3.0 to 4.6\,dex, with a step of 0.4\,dex) and \teff\,=\,12\,900\,K.}
   \label{figure2}
   \end{figure}

   \begin{figure}
   \centering
   \includegraphics[width=9cm]{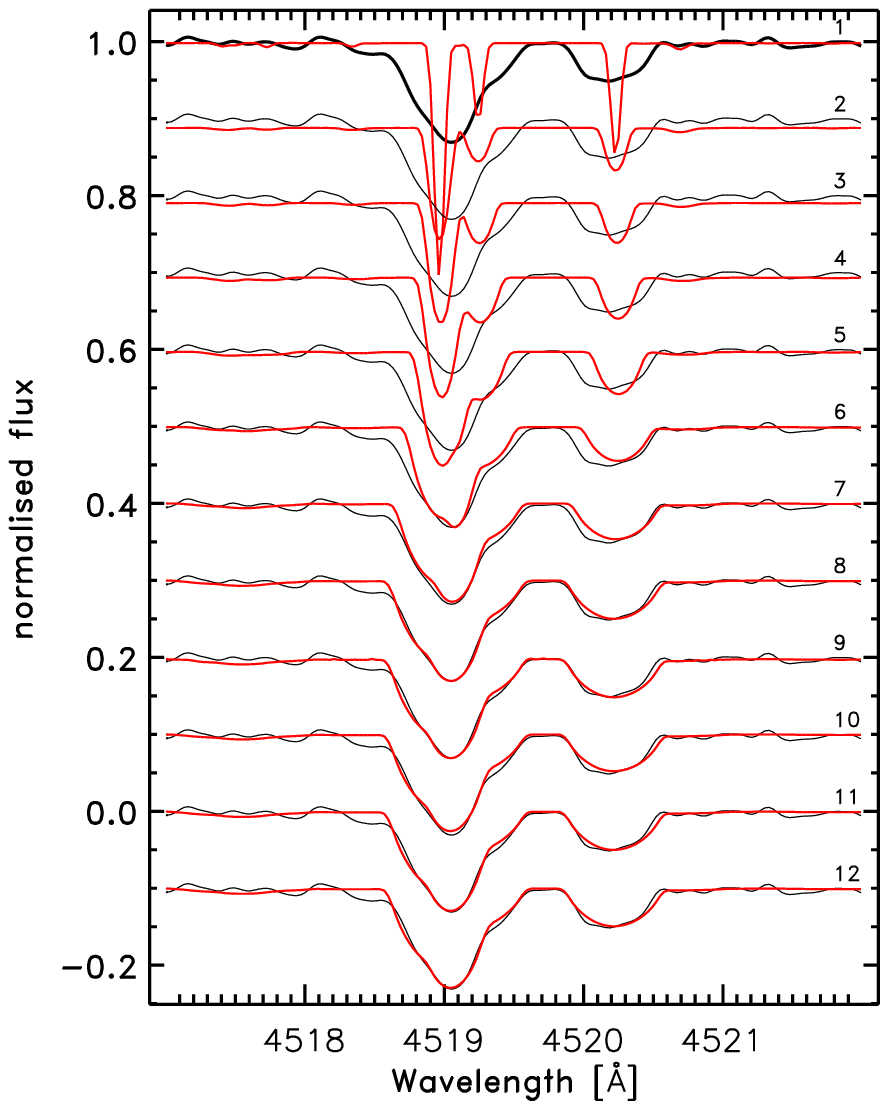}
      \caption{Comparison between the observed and synthetic spectra for the first 12 iteration steps in the case of \object{HD\,46886}. The input values are \vsini\,=\,0\,\kms, $\log\varepsilon{\rm (Fe)}$\,=\,7.50, and $\log\varepsilon{\rm (Mn)}$\,=\,7.39, whereas the output values are \vsini\,=\,23\,\kms, $\log\varepsilon{\rm (Fe)}$\,=\,6.61 and $\log\varepsilon{\rm (Mn)}$\,=\,7.29. These values are close to the final results obtained for \object{HD\,46886} on the basis of the whole spectrum (see Table\,2)}.
   \label{figure3}
   \end{figure}

   \begin{figure}
   \centering
   \includegraphics[width=9cm]{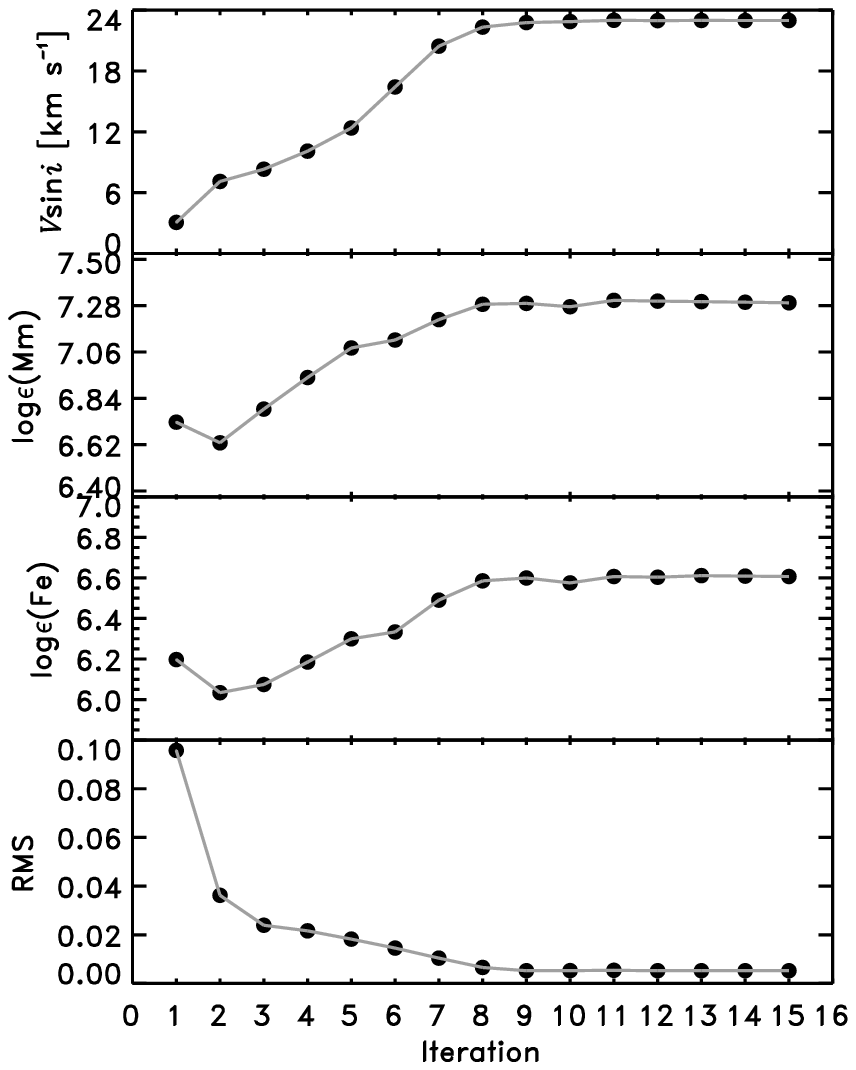}
      \caption{Variations in \vsini, $\log\varepsilon{\rm (Fe)}$, $\log\varepsilon{\rm (Mn)}$, and RMS for the case of \object{HD\,46886} and the spectral range shown in Fig.\,\ref{figure3}. }
   \label{figure4}
   \end{figure}

To perform an abundance analysis, one needs to develop an atmospheric model of the star, which requires the knowledge of its effective temperature, surface gravity and metallicity. To determine the initial stellar effective temperatures and surface gravities, we applied photometric and spectroscopic methods, respectively.

The photometric data were taken from the catalogues accessible in the GCPD database (Hauck \& Mermilliod \cite{hauck}) \footnote{http://obswww.unige.ch/gcpd/gcpd.html} and from the GAUDI archive. For every star, the average Str\"{o}mgren and Geneva photometric values were computed. 

The effective temperatures were determined on the basis of {\it uvby$\beta$} photometry using the UVBYBETA code written by Moon \& Dworetsky (\cite{moon}) and corrected by Napiwotzki et al. (\cite{napiwotzki}). For all stars for which Geneva photometry is available, we used codes and calibrations performed by K\"{u}nzli et al. (\cite{kunzli}). These calibrations can be applied to B, A, and F stars with luminosity classes V--III and are based on LTE Kurucz atmosphere models. The comparison of the \teff\, values determined from both photometric systems are shown in Fig.\,\ref{figure1}, where \teff(MN1) and \teff(MN2) refer to the effective temperatures determined from the $(b-y)_0$ and/or $c_0$ indicators, and from the [$u-b$] indicator, respectively. The most significant discrepancies between the results of Str\"{o}mgren and Geneva calibrations are found for stars with effective temperatures lower than 11\,000\,K. For a few stars with higher temperature, the differences also exceed 500\,K. This may be caused by incorrect values of photometric indices, the adopted colour excess, or the chemical peculiarity of the stars. We checked how aforementioned effects influence our \teff\, determinations. 

As mentioned above, we used Str\"{o}mgren photometry from two archives. For all the stars with both datasets available, we calculated the effective temperatures independently. The differences between the two sets of temperatures are typically lower than 200\,K. In only two cases (\object{HD\,44354} and \object{HD\,48497}), the \teff\, values obtained from the $(b-y)_0$ index differ by about 500\,K. In the case of \teff\, determined from the [$u-b$] index, all the differences are below 500\,K, typically close to 200\,K. These discrepancies are caused by incorrect values of photometric indices.

The colour excess values were determined as in Netopil et al. (\cite{netopil2009}). The colours of $UBV$ photometry were taken from the SIMBAD database. In some cases, the $UBV$ colours were unavailable. For these stars, the colour excesses $E(B-V)$ were determined from $E(b-y)$ values, calculated by the UVBYBETA code. The majority of our objects have $E(B-V)$ lower than 0.1\,mag. All the stars for which the discrepancies between \teff\, values from Str\"{o}mgren and Geneva photometry are higher than 500\,K have low colour excesses, $E(B-V) <$ 0.05\,mag. The $E(B-V)$ values are reported in Table\,1. In all cases where errors in colour indices are available, we determined the value of errors in colour excesses $E(B-V)$. In all cases, these errors are about 0.02\,mag.

Among the objects with the highest discrepancies between the temperatures determined from different photometry, only five can be chemically peculiar (\object{HD\,47759}, \object{HD\,168932}, \object{HD\,46340}, \object{HD\,168202}, \object{HD\,49713}). In Fig.\ref{figure1}, the suspected and known chemically peculiar stars discussed in Sect.\,7 are shown with different colours. As can be seen, the scatter for the sample of normal and peculiar stars is similar. The temperature calibrations of CP stars are discussed in detail by Netopil et al. (\cite{netopil2009}). If the discrepancies between \teff\, obtained from {\it uvby$\beta$} and Geneva photometry exceeded 1000\,K, we adopted the average value from the two methods presented by Napiwotzki et al. (\cite{napiwotzki}).

Once the average photometric effective temperature was obtained, the surface gravity was derived by comparing the observed Balmer profiles with theoretical ones. Four observed Balmer lines, H$\alpha$, H$\beta$, H$\gamma$, and H$\delta$, were used in the case of slowly rotating stars. Considering the uncertainties in the continuum normalisation, for rapidly rotating stars we used only H$\alpha$ and H$\beta$, because of the relatively small number of other nearby lines. The adopted average values of effective temperatures and surface gravities are shown in Table\,2. As an example, the observed H$\beta$ profile of \object{HD\,46886} is shown together with synthetic profiles obtained for different \logg\, values and \teff\,=\,12\,900\,K in Fig.\,\ref{figure2}.

For the atmospheric parameters obtained, the atmospheric models were computed with the line-blanketed LTE ATLAS9 code (Kurucz \cite{kurucz}), which treats line opacity with opacity distribution functions (ODFs). The assumed microturbulence is 2\,\kms. The synthetic spectra were computed with the SYNTHE code (Kurucz \cite{kurucz}). Both codes, ATLAS9 and SYNTHE were ported under GNU Linux by Sbordone et al. (\cite{sbordone}) and are available online \footnote{wwwuser.oat.ts.astro.it/atmos/}. The stellar line identification and the abundance analysis were performed on the basis of the VALD line lists (Kupka et al. \cite{kupka} \footnote{http://ams.astro.univie.ac.at/~vald/} and references therein). For some Mn features in the case of HgMn stars and for one Ga line (\ion{Ga}{ii} $\lambda$6334), the atomic data from Castelli \& Hubrig (\cite{castelli}) \footnote{http://wwwuser.oat.ts.astro.it/castelli/grids.html} were adopted.

\subsection{Description of the spectrum synthesis method}
Our analysis follows the methodology presented in Niemczura \& Po{\l}ubek (\cite{niemczura}) and relies on an efficient spectral synthesis based on a least squares optimisation algorithm (Takeda \cite{takeda}; Bevington \cite{bevington}). This method allows for the simultaneous determination of various parameters involved with stellar spectra and consists of the minimisation of the deviation between the theoretical flux distribution and the observed one. The synthetic spectrum depends on the stellar parameters, such as effective temperature \teff, surface gravity  \logg, rotational velocity \vsini, microturbulence \turb, radial velocity \rad, and the relative abundances of the elements \elem, where $\rm {El}$ denotes the individual element. The first two parameters and the microturbulent velocity were not determined during the iteration process but were considered as input parameters. For the majority of stars, \turb\,=\,2\,\kms\, was adopted. For only a few chemically peculiar stars, a lower value of \turb\, was assumed (\turb\,=\,0\,\kms\, for {\object{HD\,168932}, \object{HD\,179761}, \object{HD\,46886}, \object{HD\,49886}, and \turb\,=\,1\,\kms\, for \object{HD\,49481}, \object{HD\,50251}). All the other above-mentioned parameters can be determined simultaneously because they produce detectable and different spectral signatures. The \vsini\, values were determined by comparing the shapes of metal line profiles with the computed profiles, as shown in Gray\,(\cite{gray}). The limb darkening coefficient, $\epsilon$, was set to be 0.6 (see Gray\,\cite{gray}, Sect.\,18). Wade \& Rucinski (\cite{wade}) and Diaz-Cordoves et al. (\cite{diaz}) discussed the problem of linear and non-linear limb-darkening coefficients for LTE model atmospheres. According to them, for late B-type stars the linear limb-darkening coefficient is about 0.4. Assuming $\epsilon$\,=\,0.6 (typical of grey atmospheres) will have a small effect on the derived projected rotational velocities, but the differences will be comparable with the standard deviations 1\,$\sigma$. The theoretical spectrum was fitted to the normalised observed one and the continuum was objectively drawn by connecting the highest points of the analysed spectrum part. As an example of the spectral synthesis method, we show in Fig.\,\ref{figure3} a comparison between theoretical and observed spectra for the first 12 iteration steps of an illustrative spectral range of \object{HD\,46886}. In Fig.\,\ref{figure4}, variations in \vsini, $\log\varepsilon{\rm (Fe)}$, $\log\varepsilon{\rm (Mn)}$, and RMS ({\it root mean square}, the goodness-of-fit parameter) corresponding to the analysis of the same spectral part are shown. For the adopted input values, the solution is typically reached after 8 iterations. The program stops if the values of the determined parameters remain the same within 2\% for three consecutive steps. The reliability of our method is discussed in Appendix\,A. 

After estimating the atmospheric parameters \teff\, and \logg\, using the methods described above, the analysis of every star proceeded in a few steps:

\begin{enumerate}
\item Selection of the spectral parts designated for the analysis. The length of the chosen part depends mainly on the projected rotational velocity of the star. In the case of slowly and moderately rotating objects (\vsini\,$<$\,80\,\kms), short sections covering only a few spectral lines were investigated. For the stars with \vsini\,$\gtrsim$\,80\,\kms, the spectral lines are significantly blended and the continuum normalisation is difficult. For these stars, we used broader spectral ranges normalised by comparison with theoretical spectra (if necessary).

\item Line identification in the chosen part. In the case of blended lines, more than one chemical element influences the line profile. The choice of the elements to be included in the analysis proceeded in a few steps. First, we calculated the spectra individually for each element, omitting the other species. All these spectra were then normalised to the continuum level. This step was performed for every star separately, taking into account the individual stellar atmospheric parameters obtained, and the estimated rotation and radial velocity values. Second, for each element we calculated the line depths in all analysed spectral ranges. Only the most important elements producing spectral lines of significant depth were considered further.

\item For the determination of \rad, \vsini, and abundances for the adopted \teff, \logg, and \turb, the method described above was used. We adjusted the abundances to determine the closest match between the calculated and observed spectral lines.

\item The determination of mean values of \rad, \vsini, and abundances of all the chemical elements considered for a given star. These values are listed in Table\,2.

\item Estimation of errors in the obtained abundances was performed. 
\end{enumerate}

The results of the analysis are given in Table\,2. For each star, the atmospheric parameters, i.e., effective temperature, surface gravity, rotation velocity (\vsini), and the determined abundances of the elements are shown. The abundances of the chemical elements, $\log \varepsilon ({\rm {El}})$ ($\rm {El}$ denotes the individual element) are given with the standard deviations ($\sigma$) and the number of lines used in the analysis.

\section{Discussion of errors}

   \begin{figure}
   \centering
   \includegraphics[width=9cm]{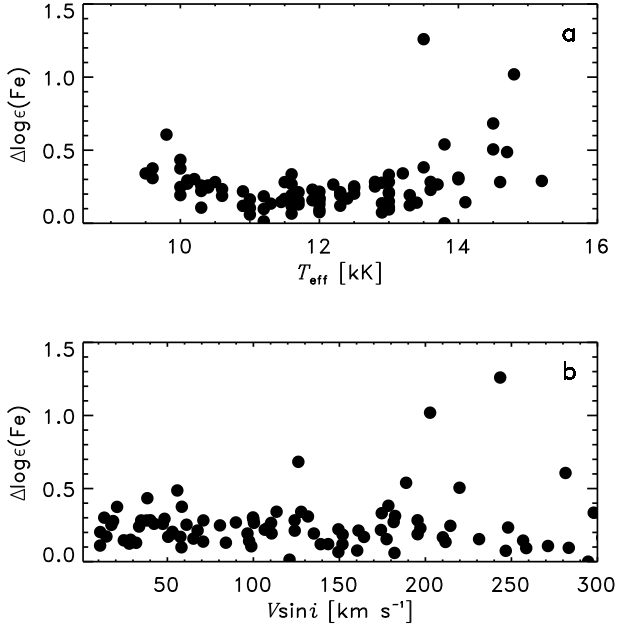}
      \caption{Uncertainties in the Fe abundances caused by a change of \teff\, by 10\%, as a function of (a) effective temperature and (b) rotation velocity.}
   \label{figure10}
   \end{figure}

   \begin{figure}
   \centering
   \includegraphics[width=9cm]{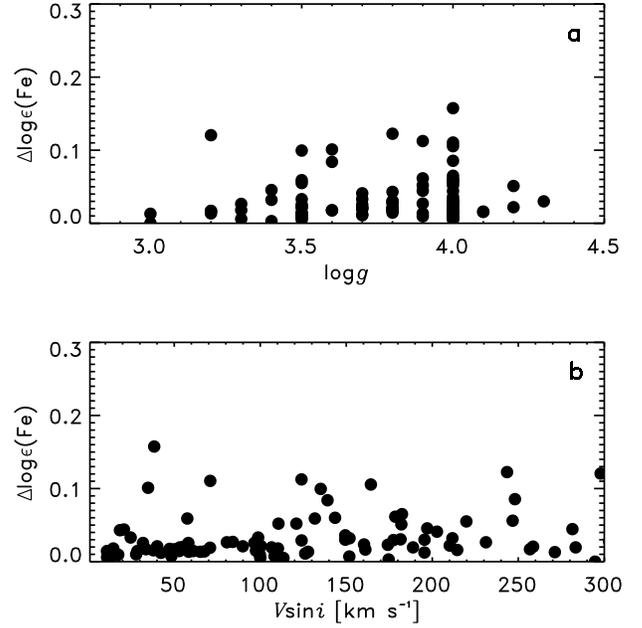}
      \caption{Uncertainties in the Fe abundances caused by a change of \logg\, by 0.1\,dex, as a function of (a) surface gravity and (b) rotation velocity.}
   \label{figure10a}
   \end{figure}

   \begin{figure}
   \centering
   \includegraphics[width=9cm]{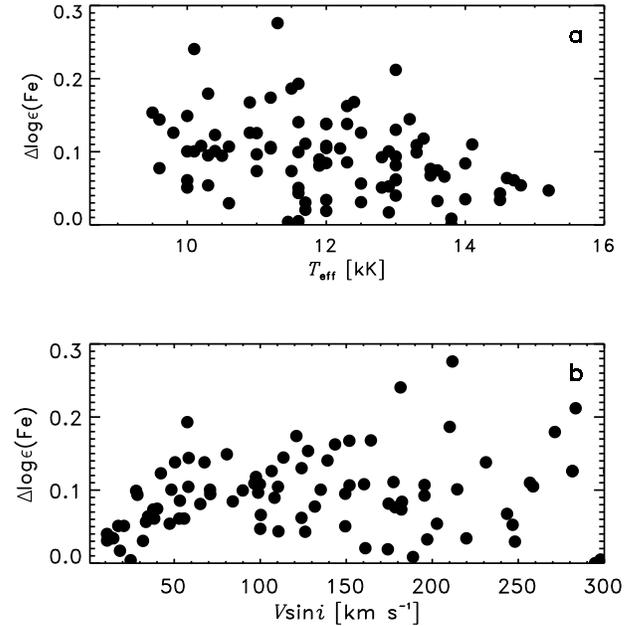}
      \caption{Uncertainties in the Fe abundances caused by a change of \turb\, by 1\,\kms, as a function of (a) effective temperature and (b) rotation velocity.}
   \label{figure10b}
   \end{figure}

Besides the line-to-line scatter, the derived abundances are affected by errors related to a number of sources of uncertainty, including the adopted atmospheric models and atomic data, most importantly the oscillator strengths. Additionally, the quality of the observed spectra and their normalisation have an influence on the chemical abundances obtained and other stellar parameters. We did not take into account non-LTE effects, which can be a significant source of errors in the case of some species, especially the light elements. For this reason, we did not use the \ion{O}{i} 7771--5\,{\AA} triplet for the determination of the oxygen abundance (Paunzen et al. \cite{paunzen1999}). The non-LTE effects for B-type stars are discussed in detail by Hempel \& Holweger (\cite{hempel2003}). They analysed the abundances of several light elements (He, C, O, Ne, Mg, Si, Ca), Fe, and two heavy elements (Sr, Ba) in 27 bright B5--B9 main-sequence stars. According to their results, departures from LTE are insignificant for Fe and Mg. For C, Ne, and Ca, the non-LTE corrections are below 0.2--0.3\,dex. The highest corrections were obtained for O (in the case of the \ion{O}{i} 7771--5\,{\AA} triplet) and for Sr. For the latter, the corrections are smaller for higher temperatures. For Si, the departures from LTE also depend on the temperature. For \teff\, lower than 13\,000\,K, the corrections are below 0.2\,dex, but for higher temperatures they can exceed 0.5\,dex. The effects of departures from LTE on nitrogen and barium are discussed by Przybilla \& Butler (\cite{przybilla_butler}) and Gigas (\cite{gigas}), and for Vega-like stars typically amount to $-$0.2 and +0.3\,dex, respectively.

The errors in the adopted stellar parameters (effective temperature, surface gravity, and microturbulence) also contribute to the uncertainties in the abundances. The systematic errors arising from the above parameters can be estimated directly by recalculating the abundances for slightly different values of \teff, \logg, and \turb\, than the adopted ones. We redetermined the mean abundances allowing for a change in \teff\, by 10\%, \logg\, by 0.1\,dex, and \turb\, by 1\,\kms. 

In Figs.\,\ref{figure10}, \ref{figure10a}, and \ref{figure10b}, the variations in the Fe abundance caused by a change in \teff, \logg, and \turb\, are shown. Iron lines are present in the spectrum of every star and for the majority of them it is the element with the highest number of lines. The uncertainties in Fe caused by errors in the effective temperature are in most cases lower than 0.3\,dex. Only for four stars (\object{HD\,176158}, \object{HD\,45760}, \object{HD\,50751}, \object{HD\,56006}), $\Delta\log\varepsilon{\rm (Fe)}$ exceeds 0.5\,dex. In the case of \object{HD\,45760}, \object{HD\,50751}, and \object{HD\,56006}, the large change in abundance is caused by the high rotational velocity of these stars (\vsini\,$>$ 200\,\kms), which can cause problems in the continuum placement. The star \object{HD\,176158} is also a rapid rotator (\vsini\,$>$\,100\,\kms), and in all these cases the Fe abundances were determined only from 1 or 2 lines. If we ignore these four objects, there is no trend between $\Delta\log\varepsilon{\rm (Fe)}$ produced by changes in \teff\, or \vsini. The sensitivity in the abundances to errors in \teff\, depends strongly on the element considered. For more discussion, see Appendix\,A. 

The variations in the Fe abundance caused by errors in the surface gravity of 0.1\,dex are in most cases lower than 0.1\,dex. There is no correlation between $\Delta\log\varepsilon{\rm (Fe)}$ and \logg. For only two stars with moderate rotation, the error in the Fe abundance is larger than 0.1\,dex (\object{HD\,168932} and \object{HD\,181761}). For all the other cases, \vsini\,$>$ 100\,\kms. There is some evidence of a small trend between $\Delta\log\epsilon{\rm (Fe)}$ and \vsini. For higher \vsini\, values, the scatter in Fig.\ref{figure10a}\,b is larger. 

The change in \turb\, by 1\,\kms\, causes variations in the Fe abundances lower than 0.2\,dex. For only three fast rotating stars (\object{HD\,171931}, \object{HD\,49643}, \object{HD\,49711}), the uncertainties in Fe slightly exceed 0.2\,dex. Here we compare only the average Fe abundances obtained from the analysis of all iron lines available in the stellar spectrum. There is a small correlation between $\Delta\log\varepsilon{\rm (Fe)}$ and \vsini\, in the case of the change in the microturbulence (see Fig.\,\ref{figure10b}). The scatter is larger for moderate and rapidly rotating stars. The impact of the uncertainties in the atmospheric parameters on our abundance results is discussed in more detail in Appendix\,A.

\section{Discussion of \vsini}
All the projected rotational velocities obtained are listed in Table\,2. We measured \vsini\, values ranging from 10 to about 300\,\kms. Low values of \vsini\, (up to 35\,\kms) were obtained for 13 stars. Moderate rotational velocities (from 40 to 80 \kms) were derived for 20 objects. The sample with high \vsini\, values is the most populated and contains 56 stars, as can be seen in the histogram presented in Fig.\,\ref{figure5}. There is a small sample of very fast rotating stars, with \vsini\, values of about 200\,\kms\, and more. The majority of stars have \vsini\, ranging from 80 to about 180\,\kms. The average \vsini\, value is 126\,\kms. Abt et al. (\cite{abt}) investigated the rotational velocities of 1092 B-type stars by determining the line width of \ion{He}{i} $\lambda$4471 and \ion{Mg}{ii} $\lambda$4481. They found the average values for B6--B8 stars of luminosity class V and IV to be 152$\pm$8 and 120$\pm$14\,\kms, respectively. For B9--B9.5 stars, the average values are 134$\pm$7 and 99$\pm$14\,\kms\, for V and IV luminosity classes, respectively. The average value obtained here for B6--B9.5 stars is consistent with the results of Abt et al. (\cite{abt}). This is a qualitative comparison because we have a much smaller sample, but both samples are expected to be affected at the same level by selection biases (e.g., the under-representation of very fast rotators). As for the confirmed CPs, the candidate CPs appear to concentrate at much lower \vsini\, values than stars without any evidence of clear abundance peculiarities (Fig.\,\ref{figure5}). A K-S test shows that the probability that the \vsini\, values of the two samples of normal and CP stars are drawn from the same distribution is only about 7\%. This is consistent with the macroscopic motions arising from rotation impeding diffusion processes (see, e.g. Abt \& Morrell \cite{abt_morrell} in the case of A stars). Furthermore, this supports the identification of our candidate CPs as bona fide members of this group.

   \begin{figure}
   \centering
   \includegraphics[width=9cm]{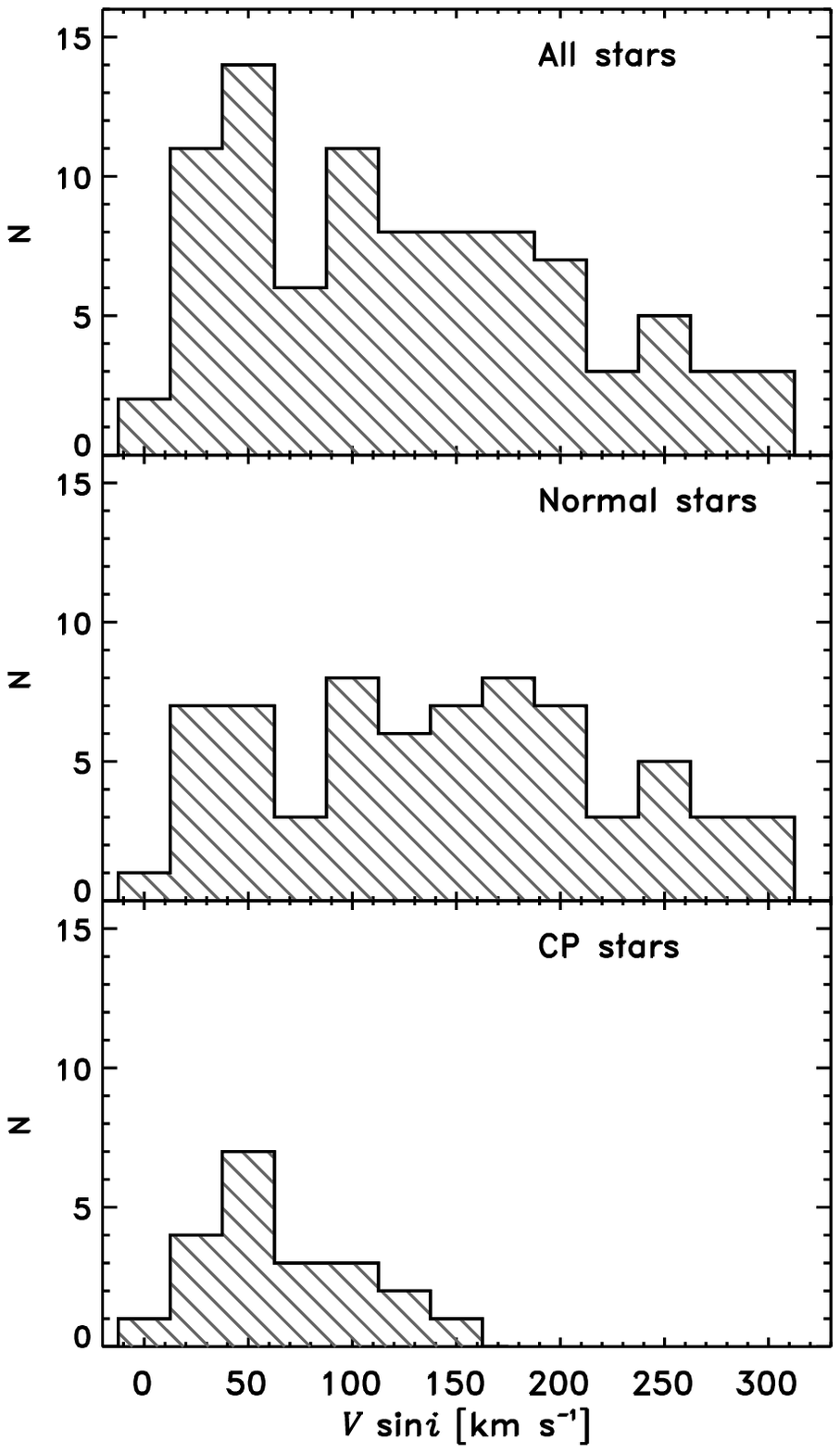}
      \caption{The distribution of rotation velocities for all analysed stars (top panel), normal (non peculiar) objects (middle panel), and chemically peculiar stars (bottom panel). For the discussion of known and newly discovered chemically peculiar stars in our sample, see Sect.\,7.
              }
         \label{figure5}
   \end{figure}

\section{Results of abundance analysis}

\begin{table}
\caption{Average abundances for the entire sample and the non-CP stars compared to the solar values (Grevesse et al.\,\cite{grevesse}).}
\begin{tabular}{l|c|c|c}
\hline
El.&$\log\varepsilon({\rm El})_{\rm all}$&$\log\varepsilon({\rm El})_{\rm normal}$& Sun\\
\hline
He  &   10.94 $\pm$   0.18&10.96 $\pm$ 0.14 & 10.93\\
C   &   8.21  $\pm$   0.41&8.21 $\pm$ 0.34  & 8.39\\
N   &   7.88  $\pm$   0.53&7.64 $\pm$ 0.17  & 7.78\\
O   &   8.72  $\pm$   0.26&8.74 $\pm$ 0.26  & 8.66\\
Ne  &   7.98  $\pm$   0.25&8.01 $\pm$ 0.26  & 7.84\\
Mg  &   7.46  $\pm$   0.47&7.56 $\pm$ 0.37  & 7.53\\
Al  &   6.02  $\pm$   0.85&6.19 $\pm$ 0.36  & 6.37\\
Si  &   7.26  $\pm$   0.41&7.22 $\pm$ 0.31  & 7.51\\
P   &   5.97  $\pm$   0.85&5.63 $\pm$ 0.37  & 5.36\\
S   &   7.06  $\pm$   0.52&7.19 $\pm$ 0.39  & 7.14\\
Ca  &   6.33  $\pm$   0.61&6.28 $\pm$ 0.55  & 6.31\\
Sc  &   3.12  $\pm$   0.37&3.08 $\pm$ 0.22  & 3.05\\
Ti  &   5.03  $\pm$   0.54&4.90 $\pm$ 0.48  & 4.90\\
V   &   4.43  $\pm$   0.22&4.38 $\pm$ 0.23  & 4.00\\
Cr  &   5.80  $\pm$   0.56&5.59 $\pm$ 0.34  & 5.64\\
Mn  &   6.34  $\pm$   1.46&5.17 $\pm$ 0.59  & 5.39\\
Fe  &   7.24  $\pm$   0.45&7.13 $\pm$ 0.29  & 7.45\\
Ni  &   6.29  $\pm$   0.58&6.23 $\pm$ 0.62  & 6.23\\
Sr  &   3.29  $\pm$   0.98&2.63 $\pm$ 0.36  & 2.92\\
Y   &   2.90  $\pm$   0.18&2.88 $\pm$ 0.21  & 2.21\\
Ba  &   2.76  $\pm$   0.52&2.64 $\pm$ 0.45  & 2.17\\
\hline 
\end{tabular}
\label{table3}
\end{table}

   \begin{figure*}
   \centering
   \includegraphics[width=18cm]{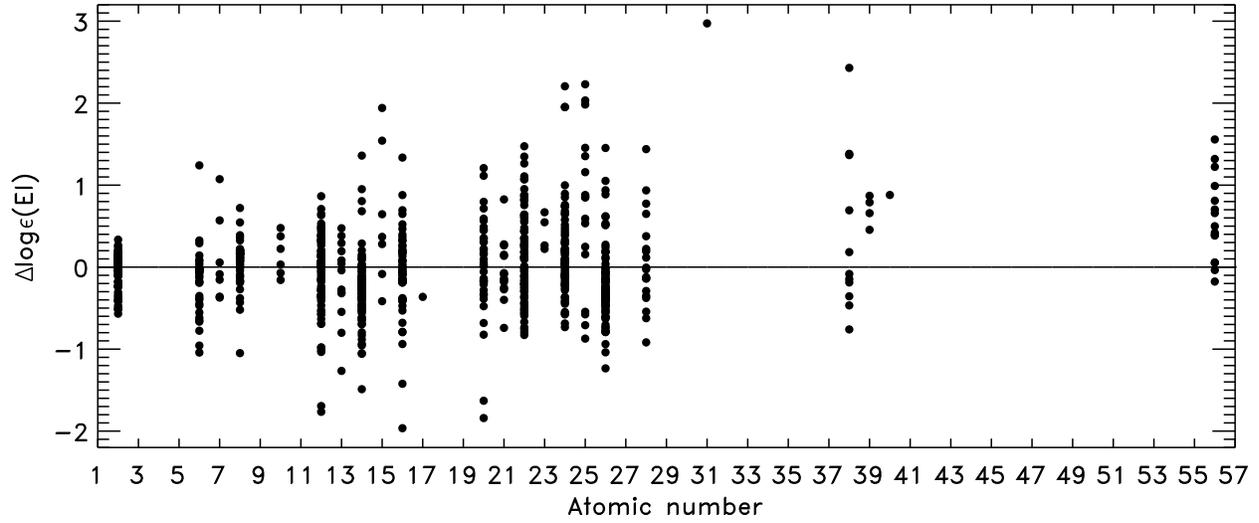}
      \caption{Obtained chemical abundances of all analysed stars compared with the solar values (Grevesse et al.\,\cite{grevesse}). The differences between stellar and solar abundances, $\Delta\log\varepsilon{\rm (El)}$, are shown as a function of the atomic number.
              }
         \label{figure6}
   \end{figure*}

   \begin{figure*}
   \centering
   \includegraphics[width=18cm]{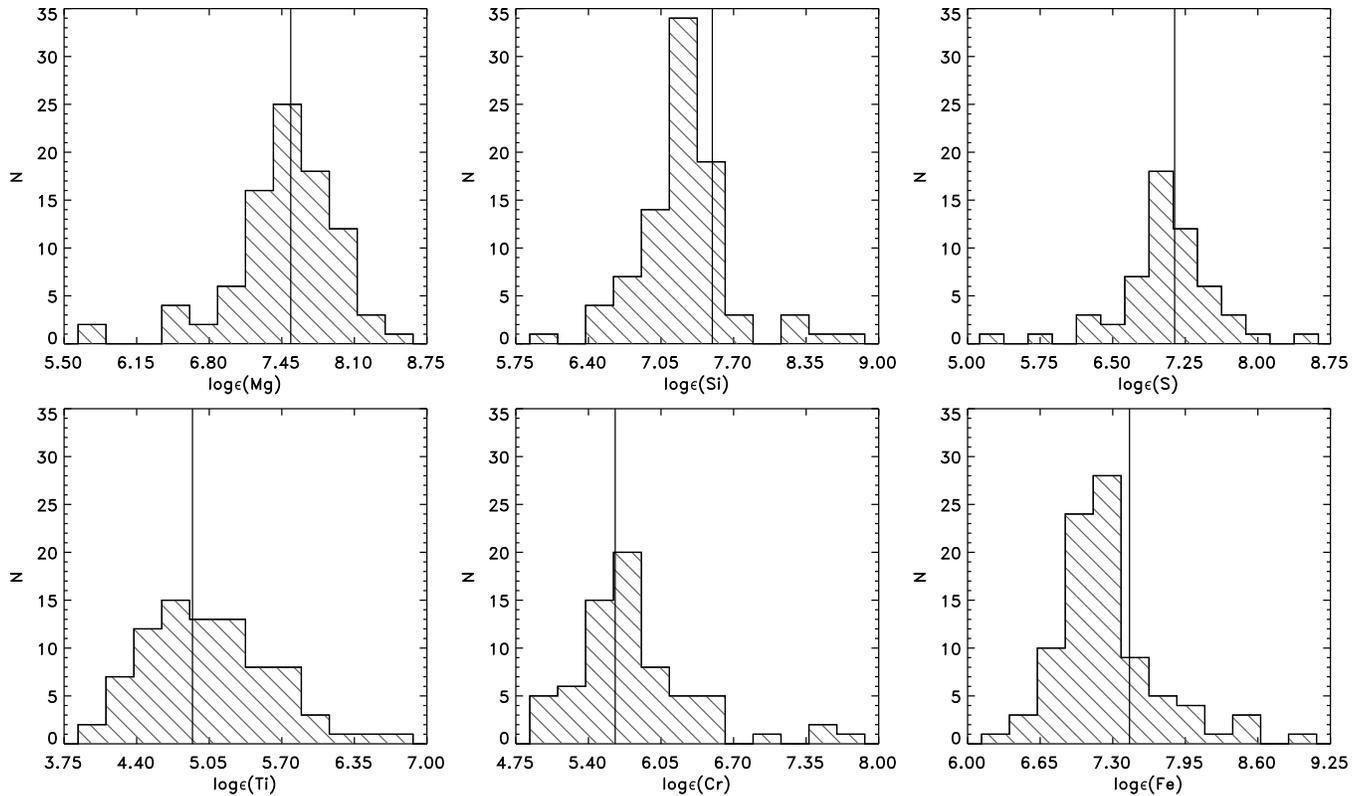}
      \caption{The distributions of the abundances of the most common elements: Mg, Si, S, Ti, Cr, and Fe. The solar abundances (Grevesse et al. \cite{grevesse}) are shown as straight lines.
              }
         \label{figure9}
   \end{figure*}

The results of the abundance analysis are presented in Table\,2. All abundances as given as $\log\varepsilon({\rm El})$, where $\log\varepsilon({\rm H})$\,=\,12 by convention. In Table\,\ref{table3}, the average abundances of the stars in our sample are compared with the solar values (Grevesse et al.\,\cite{grevesse}). Because of the neglect of non-LTE effects, which can be large for some elements, significant caution should be exercised before considering the values for the non-CP objects as representative of the present-day abundances of the ISM in the solar neighbourhood as traced by late B-type stars.

For most of the stars, the chemical abundances were determined for the first time. For the stars for which information was already available in the literature, the references to the previous investigations are also given in Table\,1. In Fig.\,\ref{figure6}, the determined chemical compositions for all stars are compared with the solar abundances of Grevesse et al. (\cite{grevesse}). From this figure, we can see that lines of iron-peak elements are clearly evident in the spectra of all stars. However, only Fe was analysed for all of them. We also investigated Sc, Ti, V, Cr, Mn, and Ni from this group. Lines of light elements are represented very sparsely in our spectra. We were able to determine the abundances of He, C, N, O, Ne, Mg, Al, Si, P, S, Cl, and Ca. We excluded the O\,I 7771--5\,{\AA} triplet from our analysis and relied instead on other O\,I lines (e.g., 5331, 6157, 6158, and 6456\,{\AA}) that are much less affected by non-LTE effects (e.g., Przybilla et al. \cite{przybilla}). Most of the abundances of these elements were determined from one or two lines only. The determination of the helium abundance may be inappropriate if the He abundance remarkably differs from the solar value. In these cases, the continuum opacities require revision and new atmospheric models are required for consistency (e.g., Jeffery et al. \cite{jeffery} \footnote{http://star.arm.ac.uk/~csj/models/Grid.html}). This may be the case for some stars in our sample. For example, He is significantly underabundant in \object{HD\,168932}, \object{HD\,44948}, \object{HD\,45153}, \object{HD\,49935}, \object{HD\,45583}, \object{HD\,46340}, \object{HD\,46886}, \object{HD\,47756}, \object{HD\,53851}, \object{HD\,55362}, and \object{HD\,56610}. For all these stars, $\log \varepsilon ({\rm He})$ is lower than 0.40. In a few cases, the He abundance is markedly higher than the solar value. This situation occurs for \object{HD\,179124}, \object{HD\,44720}, \object{HD\,47431}, \object{HD\,47759}, \object{HD\,50252}, and \object{HD\,56006}. The heavy elements (Ga, Sr, Y, Zr, Ba) were investigated only for slowly and moderately rotating stars. In most cases, only one line was available.

In Fig.\,\ref{figure9}, the histograms showing the distribution of the most common elements for the analysed late-B stars are presented. The majority of the stars have iron abundances lower than the solar value of 7.45\,dex (Grevesse et al. \cite{grevesse}). However, there are some stars with enhanced iron abundances. In the extreme case of the chemically peculiar star \object{HD\,45583}, Fe is overabundant by about 2\,dex. We should point out that if the Fe abundance is very high, then the line blanketing will be stronger. Thus, for consistency, the atmospheric models should in principle be computed with opacity distribution functions corresponding to these higher abundances. The average Fe abundance is 7.24$\pm$0.45\,dex and 7.13$\pm$0.29\,dex for the entire sample and the non-CP stars, respectively. As discussed above, only non-LTE corrections at the 0.1\,dex level are expected for Fe. The majority of the stars also have Si abundances lower than the solar value. This can be the result of a purely LTE analysis. As discussed by Hempel \& Holweger (\cite{hempel2003}), the non-LTE abundances of Si in this temperature range are higher than the LTE values. The non-LTE corrections amount to about 0.2\,dex, while the average Si abundance determined here is lower than the solar value by about 0.25\,dex. The iron-peak elements Cr and Ti have average abundances close to the solar ones. For three CP stars (\object{HD\,47759}, \object{HD\,49713}, and \object{HD\,45583}), the Cr abundances are enhanced.

In Fig.\,\ref{figure7}, the iron abundances are shown as a function of \vsini\, and \teff. There is no correlation between rotation rate and $\log\varepsilon({\rm Fe})$. On the other hand, the enhanced Fe abundances occur in most cases for stars with slow and moderate velocity. Similar dependencies were also investigated for all the other elements. A correlation between the abundances and the effective temperature was only found for Mn and Ti. In both cases, the abundances are higher with increasing \teff. These correlations can be produced because in normal, non-CP stars, the Ti and Mn lines are weaker at higher temperatures and become significantly blended with weak, unaccounted-for features of other elements (especially for rapid rotators). Therefore, we may overestimate $\log \varepsilon ({\rm Ti})$ and  $\log \varepsilon ({\rm Mn})$ for such stars. In the case of Mn, however, this correlation could have a physical cause (e.g. Jomaron et al.\,\cite{jomaron}).

   \begin{figure}
   \centering
   \includegraphics[width=9cm]{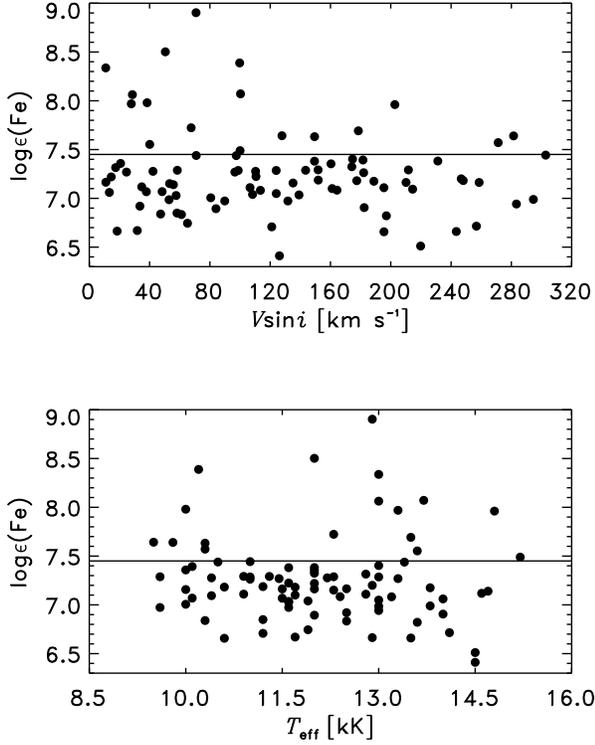}
      \caption{The obtained stellar Fe abundances, as a function of rotational velocity \vsini\, (top panel) and effective temperature (bottom panel). The horizontal line is the solar value (Grevesse et al.~\cite{grevesse}).
              }
         \label{figure7}
   \end{figure}

\section{Chemically peculiar stars}

   \begin{figure}
   \includegraphics[width=9cm]{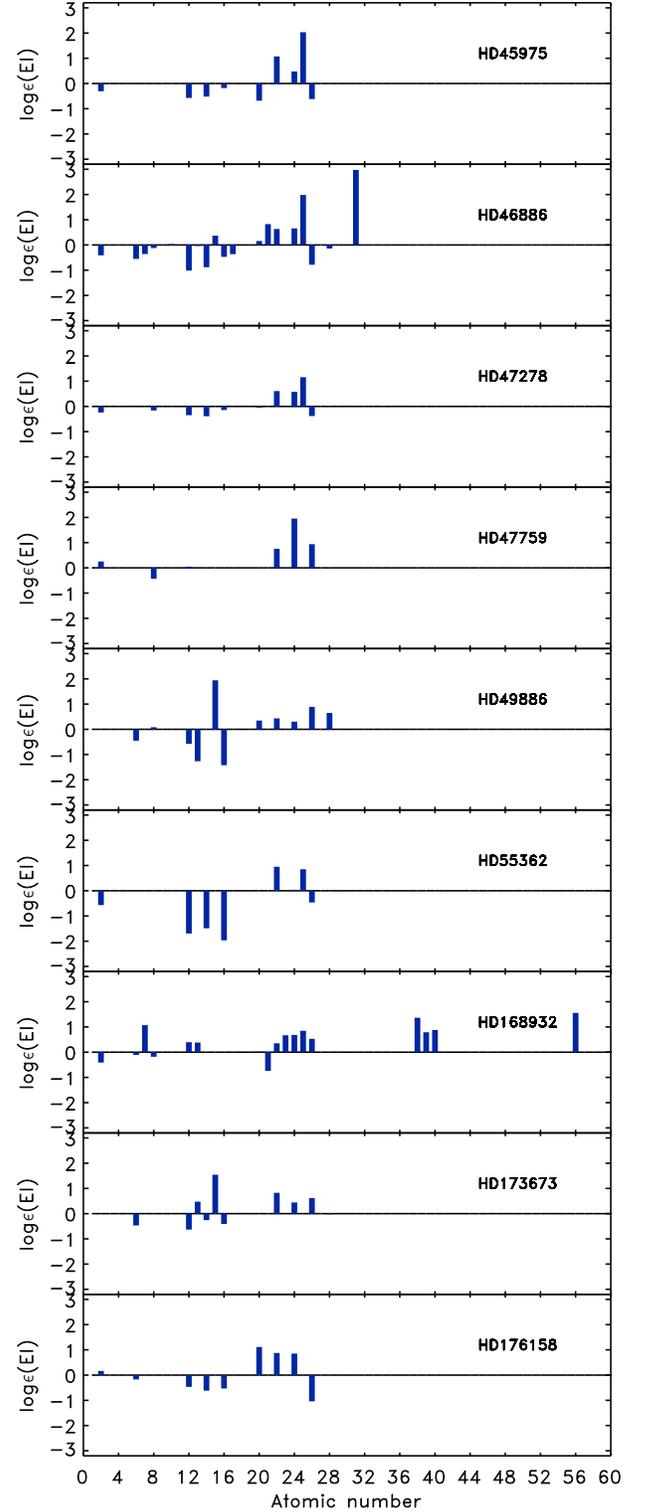}
      \caption{Elemental abundances of the new suspected chemically peculiar stars compared with the solar abundances (Grevesse et al.\,\cite{grevesse}).
              }
         \label{figure11}
   \end{figure}

Among the upper main-sequence stars in the HR diagram, there exist a number of objects with unusual surface properties, e.g., peculiar abundances of some elements. The chemically peculiar B-type stars (CP, also called Bp stars) show a variety of abundance patterns known to be associated with strong magnetic fields. Classical Bp stars have extreme overabundances of rare-earth elements and significant overabundances of Si. The latter is the most obvious anomaly in classical Bp stars. In the same range of temperatures, another group of peculiar stars exists. The mercury-manganese (HgMn) stars constitute a clearly defined sub-group of CP stars with B spectral types. They are slowly rotating, non-magnetic (i.e., a magnetic field is not definitely detected for them) and mostly young stars. The chemical peculiarities are the effect of microscopic diffusion. These stars exhibit abundance anomalies of several elements, e.g., overabundances of Hg, Mn, Y, P, Sr, and Zr, and deficiencies in He, Al, Fe, Zn, Ni. Additionally, the $\lambda$\,Bootis stars are non-magnetic, Population I dwarfs, with spectral types from late-B to early F. They show significant underabundances of metals, whereas the light elements (C, N, O, and S) have almost normal abundances compared to the Sun.

Our abundance determination of 89 late B-type stars in the {\it CoRoT} field allowed us to discover 9 new suspected CP stars and to determine for the first time abundances for 7 stars previously classified as chemically peculiar. The abundances of these objects are compared with the solar values in Figs.\,\ref{figure11} and \ref{figure12}.

Below, we provide a star-by-star description of the properties of each of those objects.

\subsection{New chemically peculiar stars}
{\it \object{HD\,45975}, NGC\,2232-6, Sp. type B9, \vsini\,=\,61$\pm$3\,\kms }. Jenkner \& Maitzen (\cite{jenkner}) searched for CP stars in the NGC\,2232 cluster. According to them, this star appears to be normal. However, they were looking only for CP2 stars (the classical magnetic peculiar stars), so they may have easily missed some other type of peculiarity. We determined a moderate rotational velocity for this star. We derived solar or close to solar abundances for all analysed light elements. On the other hand, all considered iron-peak elements besides iron are overabundant. The most enhanced abundances were obtained for Mn. This abundance pattern shows some similarities with HgMn stars, so \object{HD\,45975} can be suspected to be another member of this class.

{\it \object{HD\,46886}, Sp. type B8, \vsini\,=\,18$\pm$1\,\kms }. We determined a low rotational velocity and we were able to analyse the abundances of 19 elements. The abundances of all light elements, apart from Ne, P, and Ca, are underabundant compared to the solar values. All iron-peak elements besides Fe and Ni are enhanced. The iron abundance is lower than the solar value. The abundance of Mn is enhanced by more than 2\,dex. The only analysed heavy element, Ga, is overabundant by about 3\,dex. On the basis of this abundance pattern, we can classify \object{HD\,46886} as a HgMn star.

{\it \object{HD\,47278}, Sp. type B9, \vsini\,=\,38$\pm$2\,\kms }. We derived a moderate rotational velocity. All light elements investigated are underabundant. All analysed iron-peak elements besides Fe are overabundant. The abundance of Mn, obtained from a single line, is enhanced by about 1\,dex. Because of the small number of lines available for the analysis, we cannot conclude anything about the type of peculiarity of this object, but it could be related to HgMn stars. 

{\it \object{HD\,47759}, V\,753~Mon, Sp. type B9, \vsini\,=\,100$\pm$4\,\kms }. \object{HD\,47759} was classified as a variable star of the $\alpha^2$\,CVn-type by Kazarovets (\cite{kazarovets}). There is no information about the chemical peculiarity of this object, but $\alpha^2$\,CVn stars are rotating variables of spectral type B8p to A7p with strong magnetic fields. We found a high rotational velocity. The only light elements that we were able to analyse for this rotation velocity were He, O, and Mg. For all of them, only a single line is measurable in the spectrum. The iron-peak elements considered were Ti, Cr, and Fe. All iron-peak elements are overabundant by about 1 to 2\,dex. All elemental abundances except those for Fe were derived from 1 or 2 spectral features.

{\it \object{HD\,49886}, Sp. type B8, \vsini\,=\,11$\pm$1\,\kms }. The obtained rotational velocity is very low and allows a detailed abundance analysis to be completed. The majority of the analysed light elements are underabundant. Only O and P have enhanced abundances. The iron-peak elements have abundances close to solar values or are overabundant. The most reliable abundances were determined for iron-peak elements and two light elements, Si and P, for which more than 3 lines were analysed. 

{\it \object{HD\,55362}, Sp. type B9, \vsini\,=\,53$\pm$6\,\kms }. The abundance pattern of this star is similar to \object{HD\,47278}, and may indicate that it is another HgMn star. The abundances of the light elements are significantly lower than the solar ones. All iron-peak elements considered, apart from Fe are overabundant. The abundance of Mn is enhanced by about 1\,dex.

{\it \object{HD\,168932}, Sp. Type B9, \vsini\,=\,38$\pm$1\,\kms }. There is no determination of atmospheric parameters and chemical abundances in the literature for this star, but it was considered to be a $\lambda$\,Bootis candidate by Paunzen et al. (\cite{paunzen}). We determine a moderate rotational velocity. The abundances of the light elements differ only slightly from solar values. The most overabundant element in this group is N. All analysed iron-peak elements besides Sc are overabundant. All heavy elements are significantly overabundant, but the analysis was performed for 1 or 2 lines of each element only. This result is inconsistent with the characteristics of a typical $\lambda$\,Bootis star, but the abundance pattern of this star is peculiar.

{\it \object{HD\,173673}, Sp. type B8, \vsini\,=\,29$\pm$2\,\kms }. The moderate rotation allowed us to determine the abundances of 10 elements. The light elements turn out to be underabundant except Al, which appeared to be solar, and P, which is significantly overabundant. All analysed iron-peak elements are overabundant. Only for Fe, Ti, Si, Cr, and S are there more than 2 lines available in the spectra. The results obtained for these elements are the most reliable. The other abundances were determined from 1 or 2 features only. Considering the overabundance of P, and the deficiency in Al, we can conclude that \object{HD\,173673} is a candidate HgMn star.

{\it \object{HD\,176158}, Sp. type B9, \vsini\,=\,126$\pm$5\,\kms }. We derived a high rotational velocity for this star. The abundances of all light elements apart from He are lower than the solar values. All investigated iron-peak elements besides Fe are overabundant. The differences with the solar abundances can be about 1\,dex. Only 1 or 2 lines were available for all elements besides He.

The abundance patterns and rotational velocities of the stars listed above indicate that these objects can be chemically peculiar. We also discovered other stars with peculiarities in their spectra, but with high rotation velocities or noisy spectra. The results for them are therefore less reliable, as indicated in the following:

{\it \object{HD\,46340}, Sp. type B8, \vsini\,=\,128$\pm$4\,\kms }. We found that all elements apart from He and Sr have slightly enhanced abundances. This could be caused by binarity, as in {\it The Hipparcos and Tycho Catalogues} (Perryman 1997) this star is defined to be binary with a separation between the components $\rho\,=$\,0.650\,$\arcsec$. The same separation is given in the {\it Tycho Double Star Catalogue} (Fabricius et al. \cite{fabricius}). In the {\it All-sky Compiled Catalogue of 2.5 million stars}, Kharchenko (\cite{kharchenko}) quotes the $V$ magnitudes of both components to be 7.64 and 8.39\,mag, respectively.

{\it \object{HD\,51079}, Sp. type B8\,V, \vsini\,=\,161$\pm$8\,\kms }. We measured a high rotation velocity for this object. For this rotation value, only as many as 3 lines were available for all elements. The abundances of all elements apart from Cr are lower or close to the solar values. Chromium is overabundant by about 1\,dex.

{\it \object{HD\,53004}, Sp. type B9, \vsini\,=\,58$\pm$8\,\kms }. We obtained a moderate rotation velocity. All investigated light elements are underabundant. All iron-peak elements apart from Mn also have lower abundances than solar. Manganese is overabundant by more than 1\,dex. We classified this star as a possible HgMn object.

{\it \object{HD\,56610}, Sp. type B9, \vsini\,=\,100$\pm$10\,\kms }. We obtained a high rotational velocity. The investigated light elements are underabundant. On the other hand, all considered iron-peak elements have enhanced abundances.

{\it \object{HD\,168202}, Sp. type B9, \vsini\,=\,71$\pm$6\,\kms }. All light elements apart from Ca and Mg are underabundant. Among the iron-peak elements, only Mn is overabundant. The abundance of the rare-earth element Ba is also enhanced. Only the abundances of Ti, Cr, and Fe were derived from the analysis of more than 3 lines. If the results are reliable, \object{HD\,168202} can be classified as a suspected HgMn star.

\subsection{Known chemically peculiar stars}

   \begin{figure}
   \includegraphics[width=9cm]{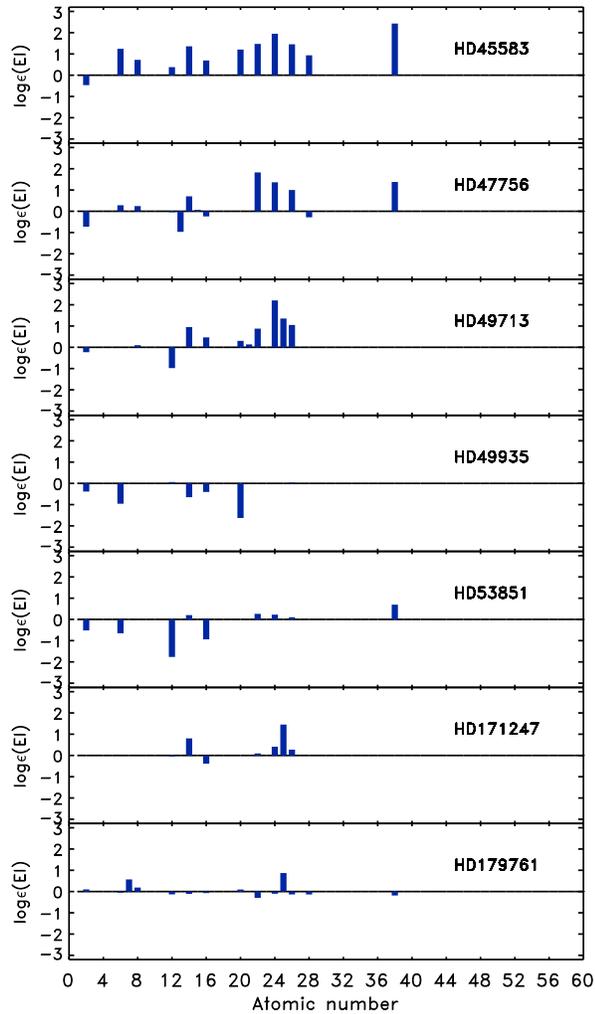}
      \caption{Elemental abundances of known chemically peculiar stars compared with the solar abundances (Grevesse et al. \cite{grevesse}).
              }
         \label{figure12}
   \end{figure}

{\it \object{HD\,45583}, V\,682~Mon, Sp. type B8\,Si, \vsini\,=\,71$\pm$6\,\kms }. This star is a member of NGC\,2232, and was found to be peculiar by Levato \& Malaroda (\cite{levato}) on the basis of spectroscopic observations. They discovered silicon lines visible at 4200, 3856, 3862, 4128, and 4130\,{\AA}. Levato \& Malaroda (\cite{levato}) also found \ion{Cr}{ii}, \ion{Fe}{ii}, and \ion{Sr}{ii} abundances to be enhanced and a great number of other faint lines present in the spectrum. Bagnulo et al. (\cite{bagnulo}) and Kudryavtsev et al. (\cite{kudryavtsev}) independently discovered a magnetic field in this star. \object{HD\,45583} is classified as a variable of the $\alpha^2$~CVn type. Two possible periods were discovered by North (\cite{north1987}). In the catalogue of Catalano \& Renson (\cite{catalano1998}), a period of 1.177\,d is given. Effective temperatures of 12\,600 and 13\,200\,K  were determined by Glagolevskij (\cite{glagolevskij}) from Q and X indices, respectively. The results of a detailed study of \object{HD\,45583} were reported by Semenko et al. (\cite{semenko}), who found a period of spectral and magnetic variability coincident with the rotation period of 1.177\,d. These authors determined \teff\,=\,13\,000\,K and \logg\,=\,4.0. According to Semenko et al. (\cite{semenko}), some elements (Fe, Si, and Cr) show a 1--2\,dex overabundance and He is underabundant by about 2\,dex with respect to the Sun.  We obtained a moderate rotational velocity and we analysed the lines of 12 elements. All light elements apart from He are overabundant. All iron-peak and heavy elements are overabundant. The most enhanced abundances were derived for Cr, Fe, and the heavy element Sr. These results are consistent with the previous ones in the literature.

{\it \object{HD\,47756}, Sp. type: B8~SiCrSr, \vsini\,=\,27$\pm$1\,\kms }. This star is a member of the NGC~2244 cluster. Cowley \& Crawford (\cite{cowley}) found \object{HD\,47756} to be peculiar on the basis of photometric criteria. Glagolevskij (\cite{glagolevskij}) obtained \teff\,=\,12\,800 and 12\,400\,K from Q and X indices, respectively. Kudryavtsev et al. (\cite{kudryavtsev}) discovered a magnetic field. We derive a low rotation velocity. Our study is the first to determine the abundance pattern of this star. All investigated light elements apart from Si are slightly underabundant. The Si abundance is close to the solar value. All the iron-peak element except Ni are overabundant. The most enhanced element is Ti. The heavy element Sr is also overabundant by about 1.5\,dex, but this value was determined for one line only. The chemical composition is typical of B8~SiSrCr stars.

{\it \object{HD\,49713}, V\,740~Mon, Sp. type B9p\,(CrEuSi), \vsini\,=\,50$\pm$2\,\kms }. \object{HD\,49713} was classified as peculiar by Walther (1949) on the basis of spectroscopic observations and was classified as a variable chemically peculiar star of the $\alpha^2$~CVn type by Kazarovets (\cite{kazarovets}). Kudryavtsev et al. (\cite{kudryavtsev}) discovered a magnetic field. We present the first determination of the atmospheric parameters and detailed chemical composition of this star. We derived a moderate rotational velocity which allows us to study the abundances of 11 elements. All investigated light elements, apart from Mg have abundances close to the solar values. All iron-peak elements are overabundant, with the exception of Sc. The most overabundant is Cr. Only the abundances of Si, Ti, Cr, and Fe were obtained on the basis of more than 3 lines. The derived chemical pattern is typical of Bp stars.

{\it \object{HD\,49935}, V741 Mon, Sp. type B8, \vsini\,=\,100$\pm$14\,\kms }. Kazarovets (\cite{kazarovets}) classified this star as a variable star of the $\alpha^2$~CVn type. All analysed light elements, including Si are underabundant. We were able to analyse only Fe of iron-peak elements. The iron abundance is close to the solar value. Only the Fe abundance was determined from more than 3 features. On the basis of these results, we cannot classify this star as chemically peculiar. 

{\it \object{HD\,53851}, Sp. type B9, \vsini\,=\,40$\pm$3\,\kms }. This star was classified as chemically peculiar by Vogt \& Faundez (\cite{vogt}) from the analysis of Str\"{o}mgren photometry. Koen \& Eyer (\cite{koen}) found \object{HD\,53851} to be variable with a period of 0.49396\,d. Beside of this, there is little information about \object{HD\,53851} in the literature. We derived a moderate rotation velocity. All light elements apart from Si are underabundant. All analysed iron-peak elements and Sr are slightly overabundant. We classify this object as a B9SiCrSr star.

{\it \object{HD\,171247}, V\,2393~Oph, Sp. type B8\,IIIsp, B8\,Si, \vsini\,=\,68$\pm$2\,\kms }. This star was classified as peculiar by Whitford (\cite{whitford}) on the basis of spectroscopic observations. Glagolevskij (\cite{glagolevskij}) determined the effective temperatures to be 12\,000 and 11\,200\,K from Q and X photometry, respectively. North \& Kroll (\cite{north1989}) derived an effective temperature of 11\,910\,K. V\,2393~Oph was claimed to be a variable star of the $\alpha^2$~CVn type. Leone et al. (\cite{leone1995}) discussed high-resolution spectroscopic observations and non-LTE calculations of the \ion{He}{i} 10\,830 {\AA} line in magnetic chemically peculiar stars, including \object{HD\,171247}. They determined \vsini\,=\,60\,\kms. Leone \& Manfre (\cite{leone1996}) performed a spectroscopic study of \object{HD\,171247} and adopted \teff\,=\,12\,166\,K and \logg\,=\,3.42. They derived the abundances with WIDTH9. The microturbulent velocity, $\xi$\,=\,2.1\,\kms, was inferred from unblended Fe lines. All unblended lines have been used to determine the rotational velocity, \vsini\,=\,50\,\kms. Leone \& Manfre (\cite{leone1996}) determined the abundances of Si, Cr, and Fe (all overabundant by 0.8\,dex). The photometric variability of this star was discovered by North (\cite{north1992}), who found the period to be 3.9124\,d, as confirmed by Catalano \& Renson (\cite{catalano1998}). Kazarovets et al. (\cite{kazarovets}) included this star in the list of variables of the $\alpha^2$~CVn type. Catanzaro et al. (\cite{catanzaro}) detected variability in \ion{He}{i} 5876\,{\AA}. They adopted \teff\,=\,11\,300\,K and \logg\,=\,3.4 from Str\"{o}mgren photometry. All elements investigated here have abundances close to the solar values or are overabundant. 

{\it \object{HD\,179761}, V\,1288\,Aql, 21\,Aql, Sp. type B8II-III, \vsini\,=\,17$\pm$1\,\kms }. This star has frequently been investigated in the past. The first determination of chemical abundances was performed by Searle et al. (\cite{searle}, see references therein), Sargent et al. (\cite{sargent}), and Durrant (\cite{durrant}). Bertaud (\cite{bertaud}) classified V\,1288\,Aql as a B8\,Si star. Kukarkin (\cite{kukarkin}) placed this star in his list of variables. Adelman (\cite{adelman1984}), Adelman \& Fuhr (\cite{adelman1985}), and Adelman (\cite{adelman1991}) performed the analysis of lines in the optical region and derived abundances close to the solar ones. Roby \& Lambert (\cite{roby}) obtained C, N, and O abundances for chemically peculiar stars, including 21\,Aql. They found C and O abundances to be in general agreement with radiative diffusion theory. Adelman et al. (\cite{adelman1993}) performed the analysis of iron-peak elements on the basis of {\it IUE} ultraviolet spectra and found good agreement between UV and optical determinations. From {\it IUE} observations, Smith (\cite{smith1993}) and Smith \& Dworetsky (\cite{smithdworetsky1993}) also obtained chemical abundances of light and iron-peak elements. They derived approximately solar abundances of Ti, Cr, Mn, Fe, and Ni. Smith (\cite{smith1994}, \cite{smith1996}, \cite{smith1997}) continued the analysis of coadded {\it IUE} spectra and investigated the ultraviolet lines of Co, Zn, Ga, and Hg. It turned out that Hg is slightly overabundant in this star. Catalano \& Renson (\cite{catalano1998}) placed V\,1288\,Aql in their list of variables as a probable B8\,Hg star. Adelman et al. (\cite{adelman2004}) analysed the lines of very heavy elements (Pt, Au, Hg, Tl, and Bi) in the UV spectra of 21\,Aql. Bychkov et al. (\cite{bychkov}) classified this star as a chemically peculiar silicon star. Ryabchikova (\cite{ryabchikova}) considered the temperature behaviour of elemental abundances in the atmosphere of magnetic peculiar stars, including 21\,Aql. For this star, Huang \& Gies (\cite{huang}) derived \logg\,=\,3.47, \teff\,=\,12\,746\,K, and a projected rotational velocity of 12$\pm$6\,\kms. Cenarro et al (\cite{cenarro}) analysed medium-resolution spectra and adopted the stellar atmospheric parameters \teff\,=\,13\,175\,K, \logg\,=\,3.27, and [Fe/H]\,=\,$-$0.14\,dex. The abundances derived here are in accordance with the previous determinations. All elements appear to be solar or slightly  underabundant. Only Mn is overabundant by about 0.5\,dex, but this is based on a single line. If the Mn abundance is reliable and taking into account the results of Smith (\cite{smith1997}), we can classify 21\,Aql as a mild HgMn star.

\section{Conclusions}
We have analysed a sample of 89 late B-type stars of spectral types B6--B9.5 located in the {\it CoRoT} field of view. All the high-resolution spectra were obtained with the FEROS and/or ELODIE spectrographs and collected in the GAUDI database, although a renormalisation of all spectra turned out to be necessary. This research is part of a global programme for deriving the elemental abundances of all B-type stars in the {\it CoRoT} field with visual magnitude below 8.0 mag and with spectra available in the GAUDI archive. In the next paper of this series, we will present the non-LTE abundance analysis of the hot B0--B5 stars (Niemczura et al., in preparation; Paper\,III).

In view of the large number of stars to be analysed, we made several basic assumptions in our study. First of all, we determined the effective temperatures of the stars from photometric calibrations and kept these fixed. Secondly, we fixed the microturbulence to a value of 2\,km\,s$^{-1}$ (or 0 and 1\,km\,s$^{-1}$ for a few stars). Thirdly, we did not allow any additional line-broadening to make the line-profile fits, i.e., we did not consider macroturbulence. Based on these assumptions, the surface gravities were derived from the comparison of observed and theoretical Balmer line profiles. For the adopted \teff, \logg, and \turb\, values, we were able to determine the chemical composition even for very rapidly rotating stars. The present study allowed us to derive the abundances of a large number of chemical species, including heavy and rare-earth elements. The dominant species in the spectrum of every star are iron-peak elements. The average Fe abundance for the entire sample is 7.24$\pm$0.45\,dex, but there are stars for which $\log\varepsilon{\rm (Fe)}$ significantly exceeds the solar value. 

As a side result, we report the discovery of at least 9 new chemically peculiar stars, including Bp and HgMn stars. Two of the most interesting objects among them are \object{HD\,46886} and \object{HD\,45975}, which are new HgMn stars with a very rich spectrum. All the other non-peculiar stars have abundance patterns that are consistent with typical abundances of B-type stars.

Our determination of the fundamental parameters and abundance patterns will serve as an important input to future seismic modelling of the stars based on {\it CoRoT} data.

\onllongtab{1}{

\begin{small}
\begin{longtable}{|l|l|l|c|c|c|c|l|l|}
\caption{The description of the analysed stars and their observations. The observation time relates to the beginning of the exposure. }\\
\hline
HD     & Sp.Type & Spectrograph & Obs. Date & Obs. Time & V    &$E(B-V)$ &Binarity & Abundance analysis \\ 
       &         &              & [UT]      & [UT]      & [mag]& [mag]   &         &                    \\ \hline

42677  & B8      & FEROS  & 2002-01-27&	03:37:43&  7.79&0.06   &                            &  \\
43406  & B9      & ELODIE & 2003-01-27&	18:40:31&  7.16&0.00   &                            &  \\
43461  & B6V     & FEROS  & 2002-01-28&	00:40:32&  6.63&0.09   &                            &  \\
43743  & B9      & FEROS  & 2002-01-28&	00:57:28&  7.70&0.00   &                            &  \\
44321  & B9      & FEROS  & 2003-01-15&	02:41:10&  7.64&0.01   &                            &  \\
44354  & B9      & FEROS  & 2003-01-17&	01:40:47&  7.70&0.01   &                            &  \\
44720  & B8      & FEROS  & 2003-01-17&	01:58:58&  7.22&0.01   &                            &  \\
44948  & B8Vp    & FEROS  & 2003-01-17&	02:05:02&  6.73&0.02   & 3, 4, 5, 6, 7, 8           &  \\
45050  & B9V     & FEROS  & 2003-01-15&	03:11:35&  6.66&0.02   & 3, 4, 5, 6, 9, 10, 11      &  \\
45153  & B8      & FEROS  & 2003-01-15&	03:16:54&  7.30&0.01   &                            &  \\
45397  & B8      & FEROS  & 2003-01-14&	03:40:40&  7.81&0.01   & 3, 5, 6                    &  \\
45515  & B8V     & FEROS  & 2003-01-17&	02:48:56&  7.88&0.01   &                            &  \\
45516  & B9      & FEROS  & 2003-01-17&	02:56:49&  7.82&0.01   &                            &  \\
45563  & B9V     & FEROS  & 2003-01-15&	03:57:36&  6.48&0.00   & 5, 6                       &  \\
45583  & B8      & FEROS  & 2003-01-17&	03:11:31&  7.98&0.01   &                            &14, 15\\
45657  & B9      & FEROS  & 2002-01-28&	02:26:30&  7.89&0.00   &                            &  \\
45709  & B9      & FEROS  & 2003-01-16&	01:05:41&  7.55&0.00   &                            &  \\
45760  & B9.5V   & FEROS  & 2003-01-15&	04:08:40&  7.56&0.00   &                            &  \\
45975  & B9      & FEROS  & 2003-01-16&	01:29:11&  7.46&0.03   &                            &  \\
46138  & B9      & FEROS  & 2003-01-14&	04:20:52&  7.46&0.00   &                            &  \\
46179  & B9V     & ELODIE & 2003-01-16&	19:26:53&  6.69&0.02   & 3, 4, 5, 6                 &  \\
46340  & B8      & FEROS  & 2002-01-28&	03:27:09&  7.65&0.00   & 3, 4, 6                    &  \\
46541  & B9      & FEROS  & 2003-01-16&	01:41:58&  7.90&0.03   &                            &  \\
       &         &        & 2003-01-18&	00:54:29&      &       &                            &  \\
46885  & B9III   & ELODIE & 2003-01-16&	20:27:27&  6.55&0.02   &                            &  \\
46886  & B9      & FEROS  & 2002-01-28&	03:43:57&  7.95&0.01   &                            &  \\
47022  & B9      & ELODIE & 2003-01-16&	22:19:10&  7.74&0.00   &                            &  \\
       &         &        & 2003-01-15&	21:11:44&      &       &                            &  \\
47221  & B9      & ELODIE & 2003-01-27&	00:00:45&  7.81&0.00   &                            &  \\
47257  & B9      & ELODIE & 2003-01-18&	19:45:23&  7.26&0.00   &                            &  \\
47272  & B9V     & ELODIE & 2003-01-24&	20:06:06&  7.62&0.03   &                            &  \\
47278  & B9      & FEROS  & 2002-01-28&	04:45:28&  7.23&0.07   &                            &  \\
47431  & B8IIIn  & ELODIE & 2003-01-16&	20:44:13&  6.57&0.03   &                            &  \\
47756  & B8IIIs  & ELODIE & 2003-01-15&	23:42:53&  6.51&0.00   &                            &  \\
47759  & B9      & ELODIE & 2003-01-22&	21:27:32&  7.52&0.00   &                            &  \\
47964  & B8III   & FEROS  & 2003-01-15&	05:17:48&  5.79&0.00   & 3, 4, 5, 6                 &  \\
48212  & B9      & ELODIE & 2003-01-26&	23:17:17&  7.80&0.04   & 4, 5, 6                    &  \\
48497  & B8      & FEROS  & 2003-01-16&	03:22:30&  7.50&0.00   &                            &  \\
48808  & B9      & FEROS  & 2003-01-14&	05:18:29&  7.45&0.04   &                            &  \\
48957  & B9      & FEROS  & 2003-01-15&	05:54:50&  8.00&0.03   & 4, 5, 6                    &  \\
49123  & B9      & FEROS  & 2003-01-17&	03:32:18&  7.22&0.02   &                            &  \\
49481  & B8      & FEROS  & 2003-01-18&	04:07:10&  6.80&0.04   &                            &  \\
49643  & B8IIIn  & FEROS  & 2003-01-16&	05:06:34&  5.75&0.00   & 3, 4, 5, 6, 9, 12          &  \\
49711  & B8V     & ELODIE & 2003-01-18&	20:52:53&  7.43&0.06   &                            &  \\
49713  & B9p     & FEROS  & 2003-01-15&	06:24:29&  7.32&0.00   &                            &  \\
49886  & B8      & FEROS  & 2003-01-17&	03:52:32&  7.58&0.00   &                            &  \\
49935  & B8      & FEROS  & 2003-01-16&	05:11:05&  6.86& --    &                            &  \\
50251  & B8V     & FEROS  & 2003-01-18&	04:25:34&  7.20&0.01   &                            &  \\
50252  & B9V     & FEROS  & 2003-01-17&	04:06:13&  7.95&0.07   &                            &  \\
50513  & B8      & FEROS  & 2003-01-17&	04:21:58&  7.82&0.00   &                            &  \\
50751  & B8      & FEROS  & 2003-01-17&	04:37:56&  7.90&0.00   &                            &  \\
51079  & B8V     & FEROS  & 2003-01-18&	05:12:03&  7.94&0.04   &                            &  \\
52312  & B9III   & FEROS  & 2003-01-18&	05:35:03&  5.96&0.01   & 3, 4, 5, 6                 &  \\
53004  & B9      & FEROS  & 2003-01-14&	07:00:14&  7.26&0.03   & 3, 4, 5, 6                 &  \\
53083  & B8      & FEROS  & 2003-01-15&	07:24:17&  7.10&0.02   &                            &  \\
53851  & B9      & FEROS  & 2003-01-14&	07:21:06&  7.58&0.00   &                            &  \\
54929  & B9      & FEROS  & 2003-01-18&	06:42:53&  7.34&0.02   & 3, 4, 5, 6                 &  \\
55362  & B9      & FEROS  & 2003-01-14&	07:58:01&  7.95&0.00   &                            &  \\
55793  & B8      & FEROS  & 2003-01-14&	08:07:45&  7.86&0.01   &                            &  \\
56006  & B8      & FEROS  & 2003-01-17&	07:04:32&  7.69&0.06   &                            &  \\
56446  & B8III   & FEROS  & 2003-01-15&	08:11:38&  6.65&0.00   &                            &  \\
56610  & B9      & FEROS  & 2003-01-17&	07:19:19&  7.80&0.00   &                            &  \\
56613  & B9      & FEROS  & 2003-01-18&	07:24:51&  7.10&0.00   &                            &  \\
57275  & B9      & ELODIE & 2001-11-28&	03:24:06&  6.90&0.00   & 3, 4, 6, 5, 7              &  \\
57293  & B9III-IV& FEROS  & 2003-01-16&	07:54:23&  7.99&0.03   & 3, 4, 6, 5, 7, 12          &  \\
168202 & B9      & ELODIE & 2002-08-15&	19:37:44&  7.50&0.00   &                            &  \\
168932 & B9      & FEROS  & 2001-07-05&	02:56:50&  7.28&0.00   & 1, 2                       &  \\
169224 & B9      & ELODIE & 2002-08-16&	20:11:21&  7.50&0.04   &                            &  \\
169225 & B9      & FEROS  & 2001-07-05&	03:06:33&  7.72&0.01   &                            &  \\
169512 & B9      & ELODIE & 2002-08-19&	19:36:33&  7.91&0.15   & 3, 4, 5, 6, 7              &  \\
169578 & B9V     & FEROS  & 2001-07-06&	07:28:52&  6.73&0.09   & 1, 2                       &  \\
170783 & B5      & ELODIE & 2000-06-18&	00:38:18&  7.73&0.34   &                            &  \\
       &         &        & 2000-06-18&	00:57:09&      &       &                            &  \\
170935 & B8      & ELODIE & 2002-08-18&	19:27:06&  7.38&0.20   &                            &  \\
171247 & B8IIIsp & FEROS  & 2001-07-07&	04:42:34&  6.42&0.07   & 3, 4, 5, 6                 &13\\
171931 & B9      & FEROS  & 2005-06-18&	02:07:17&  9.19&0.26   &                            &  \\
172850 & B9      & FEROS  & 2001-07-05&	04:10:15&  7.75&0.08   &                            &  \\
173673 & B8      & FEROS  & 2001-07-05&	04:43:37&  7.65&0.13   &                            &  \\
174701 & B9      & FEROS  & 2001-07-05&	06:07:22&  8.00&0.15   & 4, 5, 6                    &  \\
174836 & B9      & FEROS  & 2001-07-05&	06:22:37&  7.89&0.00   &                            &  \\
174884 & B8      & ELODIE & 2000-06-10&	01:24:45&  7.99&0.00   &                            &  \\
       &         &        & 2000-06-10&	02:06:56&      &       &                            &  \\
176076 & B9      & FEROS  & 2001-07-05&	07:19:25&  7.28&0.06   &                            &  \\
176158 & B9      & FEROS  & 2001-07-05&	07:39:42&  7.52&0.22   &                            &  \\
176258 & B9V     & FEROS  & 2001-07-07&	05:44:09&  7.52&0.00   &                            &  \\
177756 & B9Vn    & FEROS  & 2001-07-06&	02:32:47&  3.43&0.00   &                            &  \\
177880 & B5V     & FEROS  & 2001-07-06&	04:23:07&  6.76&0.26   & 3, 4, 5, 6                 &  \\
178744 & B5Vn    & FEROS  & 2001-07-05&	09:28:56&  6.33&0.00   &                            &  \\
179124 & B9V     & FEROS  & 2001-07-06&	09:12:42&  6.90&0.04   &                            &  \\
179761 & B8II-III& FEROS  & 2001-07-07&	06:03:24&  5.14&0.05   & 3, 4, 5, 6                 &16, 17, 18, 19, 20, 21, 22,\\
       &         &        &           &         &      &       &                            &23, 24, 25, 26, 27, 28, 29,\\
       &         &        &           &         &      &       &                            &30, 31 \\
181440 & B9III   & FEROS  & 2001-07-05&	09:44:56&  5.48&0.02   &                            &  \\
181761 & B8      & FEROS  & 2001-07-04&	09:34:12&  7.94&0.01   &                            &  \\
182198 & B9V     & FEROS  & 2001-07-06&	04:35:39&  7.94&0.04   &                            &  \\
\hline
\end{longtable}
References: (1) Frankowski et al. (2007); (2) Makarov et al. (2005); (3) Dommanget \& Nys (2000); (4) Dommanget \& Nys (2002); (5) Fabricius et al. (2002); (6) Mason et al. (2001); (7) Abt (1985); (8) Horch et al. (2006); (9) Sowell \& Wilson (1993); (10) Abt \& Boonyarak (2004); (11) Abt (2005); (12) Hartkopf et al. (2000); (13) Leone \& Manfre (1996); (14) Semenko et al. (2008); (15) Levato \& Malaroda (1974); (16) Searle et al. (1966); (17) Sargent et al. (1969); (18) Durrant (1970); (19) Adelman (1984); (20) Adelman \& Fuhr (1985); (21) Adelman (1991); (22) Roby \& Lambert (1990); (23) Adelman et al. (1993); (24) Smith (1993); (25) Smith \& Dworetsky (1993); (26) Smith (1994); (27) Smith (1996); (28) Smith (1997); (29) Adelman et al. (2004); (30) Ryabchikova  (2005); (31) Cenarro et al. (2007).
\end{small}
}
\onllongtabL{2}{
\begin{landscape}
\begin{scriptsize}
\begin{longtable}{|c|c|c|c|c|c|c|c|c|c|c|c|c|c|c|c|c|c|c|c|c|c|c|c|c|c|c|c|}
\caption{The atmospheric parameters, abundances of chemical elements and rotation velocities of all analysed stars.}\\
\hline
HD&$T_{\rm eff}$&$\log g$&$V\sin i$&He&C&N&O&Ne&Mg  &Al  &Si  &P   &S   &Cl  &Ca  &Sc  &Ti  &V   &Cr  &Mn  &Fe  &Ni  &Ga  &Sr  &Y   &Zr  &Ba\\ 
  &[K]          &    &{km~s}$^{-1}$&  & & & &  &    &    &    &    &    &    &    &    &    &    &    &    &    &    &    &    &    &    &  \\ 
\hline \hline
  42677&10000&3.5&135& -    &7.35& -  &8.82& -  &7.66& -  &7.44& -  & -  & -  &6.90& -  &4.31& -  &5.50& -  &7.16& -  & -  & -  & -  & -  &1.99\\
       &     &   &  4& -    & -  & -  & -  & -  &0.04& -  &0.07& -  & -  & -  & -  & -  &0.16& -  &0.04& -  &0.14& -  & -  & -  & -  & -  &0.21\\
       &     &   &   & -    &(1) & -  &(1) & -  &(2) & -  &(2) & -  & -  & -  &(1) & -  &(4) & -  &(2) & -  &(9) & -  & -  & -  & -  & -  &(2) \\ \hline
  43406&11600&3.6&139&10.93 & -  & -  & -  & -  &7.89& -  &7.00& -  &7.75& -  & -  & -  &5.03& -  &5.73& -  &7.04& -  & -  & -  & -  & -  & -  \\
       &     &   & 11& -    & -  & -  & -  & -  &0.07& -  &0.11& -  & -  & -  & -  & -  & -  & -  & -  & -  &0.19& -  & -  & -  & -  & -  & -  \\
       &     &   &   &(1)   & -  & -  & -  & -  &(3) & -  &(2) & -  &(1) & -  & -  & -  &(1) & -  &(1) & -  &(5) & -  & -  & -  & -  & -  & -  \\ \hline
  43461&13600&3.4&197&11.13 &8.31& -  & -  & -  &7.39& -  &7.39& -  &6.58& -  &6.46& -  &5.97& -  & -  & -  &6.82&5.90& -  & -  & -  & -  & -  \\
       &     &   & 14&0.06  &0.29& -  & -  & -  &0.25& -  & -  & -  &0.08& -  & -  & -  & -  & -  & -  & -  &0.08& -  & -  & -  & -  & -  & -  \\
       &     &   &   &(7)   &(2) & -  & -  & -  &(2) & -  &(1) & -  &(2) & -  &(1) & -  &(1) & -  & -  & -  &(4) &(1) & -  & -  & -  & -  & -  \\ \hline
  43743&11000&4.2&182&11.06 & -  & -  &8.83& -  &7.66& -  &7.25& -  &7.09& -  &6.48&3.03&4.85& -  &5.84& -  &7.26& -  & -  & -  & -  & -  &2.87\\
       &     &   & 10& -    & -  & -  &0.01& -  &0.08& -  &0.11& -  & -  & -  & -  & -  &0.09& -  &0.08& -  &0.13& -  & -  & -  & -  & -  & -  \\
       &     &   &   &(1)   & -  & -  &(2) & -  &(4) & -  &(2) & -  &(1) & -  &(1) &(1) &(5) & -  &(3) & -  &(11)& -  & -  & -  & -  & -  &(1) \\ \hline
  44321&11000&3.7& 99&11.11 & -  & -  &8.85& -  &7.38&6.06&7.39& -  & -  & -  &6.03& -  &4.74& -  &5.54& -  &7.29& -  & -  & -  & -  & -  & -  \\
       &     &   &  6&0.01  & -  & -  &0.01& -  &0.16& -  &0.06& -  & -  & -  & -  & -  &0.14& -  &0.06& -  &0.11& -  & -  & -  & -  & -  & -  \\
       &     &   &   &(2)   & -  & -  &(2) & -  &(4) &(1) &(4) & -  & -  & -  &(1) & -  &(8) & -  &(3) & -  &(20)& -  & -  & -  & -  & -  & -  \\ \hline
  44354&13000&3.9&124&11.18 &8.33& -  &8.88& -  &7.86& -  &7.31& -  &7.25& -  & -  &3.47&5.34& -  &5.84& -  &7.00& -  & -  & -  & -  & -  & -  \\
       &     &   &  8&0.06  & -  & -  & -  & -  &0.20& -  &0.09& -  &0.11& -  & -  & -  &0.01& -  &0.52& -  &0.11& -  & -  & -  & -  & -  & -  \\
       &     &   &   &(7)   &(1) & -  &(1) & -  &(2) & -  &(3) & -  &(4) & -  & -  &(1) &(2) & -  &(2) & -  &(6) & -  & -  & -  & -  & -  & -  \\ \hline
  44720&14000&4.0&182&11.16 &8.43& -  &8.83& -  &7.43& -  &7.15& -  &7.50& -  & -  & -  &6.16& -  & -  & -  &6.91& -  & -  & -  & -  & -  & -  \\
       &     &   & 15&0.05  & -  & -  & -  & -  &0.57& -  &0.18& -  &0.19& -  & -  & -  & -  & -  & -  & -  &0.14& -  & -  & -  & -  & -  & -  \\
       &     &   &   &(7)   &(1) & -  &(1) & -  &(2) & -  &(3) & -  &(3) & -  & -  & -  &(1) & -  & -  & -  &(3) & -  & -  & -  & -  & -  & -  \\ \hline
  44948&13400&4.0& 98&10.63 & -  & -  & -  & -  &6.84&5.57&8.19& -  &7.04& -  & -  & -  &5.97& -  &6.02& -  &7.44& -  & -  & -  & -  & -  & -  \\
       &     &   &  7&0.07  & -  & -  & -  & -  & -  & -  &0.18& -  & -  & -  & -  & -  & -  & -  & -  & -  &0.16& -  & -  & -  & -  & -  & -  \\
       &     &   &   &(2)   & -  & -  & -  & -  &(1) &(1) &(8) & -  &(1) & -  & -  & -  &(1) & -  & -  & -  &(6) & -  & -  & -  & -  & -  & -  \\ \hline
  45050&11600&4.0&149&11.04 & -  & -  &8.84& -  &7.56& -  &7.13& -  &7.46& -  &6.32& -  &5.38& -  &6.28& -  &7.38& -  & -  & -  & -  & -  & -  \\
       &     &   &  7&0.03  & -  & -  &0.07& -  &0.38& -  &0.09& -  & -  & -  & -  & -  &0.04& -  &0.08& -  &0.15& -  & -  & -  & -  & -  & -  \\
       &     &   &   &(2)   & -  & -  &(2) & -  &(2) & -  &(2) & -  &(1) & -  &(1) & -  &(4) & -  &(2) & -  &(9) & -  & -  & -  & -  & -  & -  \\ \hline
  45153&12300&4.0&143&10.46 & -  & -  & -  & -  &7.46& -  &7.13& -  &7.32& -  &6.13& -  &5.00& -  & -  & -  &7.29& -  & -  & -  & -  & -  & -  \\
       &     &   &  5&0.05  & -  & -  & -  & -  &0.24& -  &0.08& -  & -  & -  & -  & -  &0.19& -  & -  & -  &0.18& -  & -  & -  & -  & -  & -  \\
       &     &   &   &(3)   & -  & -  & -  & -  &(2) & -  &(2) & -  &(1) & -  &(1) & -  &(3) & -  & -  & -  &(7) & -  & -  & -  & -  & -  & -  \\ \hline
  45397&12400&3.9&164&11.10 & -  & -  &8.83& -  &7.77& -  &7.03& -  &7.33& -  &7.04& -  &4.77& -  &5.68& -  &7.03& -  & -  & -  & -  & -  & -  \\
       &     &   &  7&0.08  & -  & -  &0.25& -  &0.06& -  &0.10& -  & -  & -  & -  & -  &0.04& -  & -  & -  &0.23& -  & -  & -  & -  & -  & -  \\
       &     &   &   &(2)   & -  & -  &(2) & -  &(2) & -  &(2) & -  &(1) & -  &(1) & -  &(2) & -  &(1) & -  &(7) & -  & -  & -  & -  & -  & -  \\ \hline
  45515&10600&4.0&248&10.99 & -  & -  &8.99& -  &7.55& -  &7.19& -  & -  & -  & -  & -  &4.38& -  & -  & -  &7.18& -  & -  & -  & -  & -  & -  \\
       &     &   & 24&      & -  & -  & -  & -  &0.18& -  & -  & -  & -  & -  & -  & -  &0.15& -  & -  & -  &0.25& -  & -  & -  & -  & -  & -  \\
       &     &   &   &(1)   & -  & -  &(1) & -  &(2) & -  &(1) & -  & -  & -  & -  & -  &(2) & -  & -  & -  &(4) & -  & -  & -  & -  & -  & -  \\ \hline
  45516&12900&4.0&247&11.12 &8.68& -  & -  & -  &7.53& -  &7.27& -  &7.41& -  & -  & -  &5.54& -  &6.15& -  &7.20& -  & -  & -  & -  & -  & -  \\
       &     &   & 19&0.02  & -  & -  & -  & -  &0.50& -  & -  & -  &0.11& -  & -  & -  &0.52& -  & -  & -  &0.13& -  & -  & -  & -  & -  & -  \\
       &     &   &   &(2)   &(1) & -  & -  & -  &(2) & -  &(1) & -  &(3) & -  & -  & -  &(2) & -  &(1) & -  &(3) & -  & -  & -  & -  & -  & -  \\ \hline
  45563&10900&4.0&107&11.10 & -  & -  &8.83& -  &7.66& -  &7.37& -  &7.66& -  & -  & -  &4.65& -  &5.73& -  &7.11& -  & -  & -  & -  & -  & -  \\
       &     &   &  5&0.02  & -  & -  & -  & -  &0.01& -  &0.05& -  & -  & -  & -  & -  &0.08& -  &0.12& -  &0.18& -  & -  & -  & -  & -  & -  \\
       &     &   &   &(2)   & -  & -  &(1) & -  &(2) & -  &(3) & -  &(1) & -  & -  & -  &(4) & -  &(3) & -  &(15)& -  & -  & -  & -  & -  & -  \\ \hline
  45583&12900&3.7& 71&10.33 &9.61& -  &9.39& -  &7.92& -  &8.87& -  &7.81& -  &7.52& -  &6.44& -  &7.59& -  &8.92&7.17& -  &5.35& -  & -  & -  \\
       &     &   &  6&0.17  & -  & -  &0.33& -  &0.04& -  &0.08& -  &0.28& -  & -  & -  &0.16& -  &0.16& -  &0.19&0.13& -  & -  & -  & -  & -  \\
       &     &   &   &(5)   &(1) & -  &(2) & -  &(3) & -  &(11)& -  &(6) & -  &(1) & -  &(7) & -  &(16)& -  &(51)&(2) & -  &(1) & -  & -  & -  \\ \hline
  45657&10400&4.1&214&10.98 & -  & -  &8.81& -  &7.24& -  &7.10& -  & -  & -  & -  & -  &4.07& -  &5.52& -  &7.09& -  & -  & -  & -  & -  & -  \\
       &     &   &  5&  -   & -  & -  & -  & -  &0.24& -  &0.01& -  & -  & -  & -  & -  &0.28& -  & -  & -  &0.27& -  & -  & -  & -  & -  & -  \\
       &     &   &   &(1)   & -  & -  &(1) & -  &(2) & -  &(2) & -  & -  & -  & -  & -  &(2) & -  & -  & -  &(2) & -  & -  & -  & -  & -  & -  \\ \hline
  45709&11500&4.2&210&11.07 & -  & -  &8.74& -  &7.71& -  &7.15& -  & -  & -  & -  & -  &5.25& -  &5.86& -  &7.16& -  & -  & -  & -  & -  & -  \\
       &     &   &  7&  -   & -  & -  & -  & -  &0.25& -  &0.17& -  & -  & -  & -  & -  &0.24& -  & -  & -  &0.18& -  & -  & -  & -  & -  & -  \\
       &     &   &   &(1)   & -  & -  &(1) & -  &(2) & -  &(3) & -  & -  & -  & -  & -  &(2) & -  &(1) & -  &(5) & -  & -  & -  & -  & -  & -  \\ \hline
  45760& 9800&3.9&281&10.87 & -  & -  &8.88& -  &7.85& -  &7.41& -  & -  & -  &6.62& -  &5.14& -  & -  & -  &7.64& -  & -  & -  & -  & -  & -  \\
       &     &   &  7&  -   & -  & -  & -  & -  &0.01& -  &0.26& -  & -  & -  & -  & -  &0.02& -  & -  & -  &0.09& -  & -  & -  & -  & -  & -  \\
       &     &   &   &(1)   & -  & -  &(1) & -  &(2) & -  &(2) & -  & -  & -  &(1) & -  &(2) & -  & -  & -  &(3) & -  & -  & -  & -  & -  & -  \\ \hline
  45975&12500&4.0& 61&10.66 &8.38 & -  & -  & -  &6.96& -  &6.99& -  &6.96& -  &5.63& -  &5.97& -  &6.12&7.42&6.83& -  & -  & -  & -  & -  & -  \\
       &     &   &  3&0.05  &0.78 & -  & -  & -  &0.28& -  &0.09& -  &0.13& -  & -  & -  &0.14& -  &0.07&0.13&0.17& -  & -  & -  & -  & -  & -  \\
       &     &   &   &(3)   &(2)  & -  & -  & -  &(2) & -  &(6) & -  &(2) & -  &(1) & -  &(7) & -  &(3) &(10)&(13)& -  & -  & -  & -  & -  & -  \\ \hline
  46138&11600&3.9&111&11.10 & -  & -  &8.80& -  &7.35& -  &7.33& -  &7.28& -  & -  & -  &4.80& -  &5.39& -  &7.22& -  & -  & -  & -  & -  & -  \\
       &     &   &  7&0.03  & -  & -  &0.02& -  &0.14& -  &0.11& -  &0.26& -  & -  & -  &0.01& -  & -  & -  &0.13& -  & -  & -  & -  & -  & -  \\
       &     &   &   &(3)   & -  & -  &(2) & -  &(2) & -  &(5) & -  &(2) & -  & -  & -  &(2) & -  &(1) & -  &(6) & -  & -  & -  & -  & -  & -  \\ \hline
  46179&10900&4.0&152&11.04 & -  & -  &8.85& -  &7.58& -  &7.60& -  & -  & -  & -  & -  &4.61& -  &5.79& -  &7.29& -  & -  & -  & -  & -  & -  \\
       &     &   & 13&0.02  & -  & -  & -  & -  &0.02& -  &0.04& -  & -  & -  & -  & -  &0.09& -  &0.18& -  &0.18& -  & -  & -  & -  & -  & -  \\
       &     &   &   &(2)   & -  & -  &(1) & -  &(2) & -  &(2) & -  & -  & -  & -  & -  &(3) & -  &(2) & -  &(9) & -  & -  & -  & -  & -  & -  \\ \hline
  46340& 9500&3.9&128&10.61 & -  & -  &8.83& -  &8.02& -  &7.61& -  & -  & -  &6.77&3.43&5.26&4.27&5.97&5.55&7.64& -  & -  &2.77& -  & -  &2.67\\
       &     &   &  4&0.02  & -  & -  &0.01& -  &0.16& -  &0.10& -  & -  & -  &0.19&0.11&0.15& -  &0.07& -  &0.09& -  & -  & -  & -  & -  & -  \\
       &     &   &   &(2)   & -  & -  &(2) & -  &(5) & -  &(2) & -  & -  & -  &(2) &(2) &(13)&(1) &(4) &(1) &(16)& -  & -  &(1) & -  & -  &(1) \\ \hline
  46541& 9600&3.5& 58&  -   &7.99& -  &8.73& -  &7.52&6.45&7.38& -  & -  & -  &5.47&2.90&4.79& -  &5.70&4.81&7.29& -  & -  & -  & -  & -  &2.56\\
       &     &   &  1&  -   &0.57& -  &0.01& -  &0.05& -  &0.11& -  & -  & -  &0.75&0.08&0.06& -  &0.08& -  &0.08& -  & -  & -  & -  & -  &0.11\\
       &     &   &   &      &(2) & -  &(2) & -  &(4) &(1) &(5) & -  & -  & -  &(2) &(4) &(17)& -  &(12)&(1) &(34)& -  & -  & -  & -  & -  &(2) \\ \hline
  46885&11700&3.6& 32&10.86 & -  & -  &8.70& -  &6.49& -  &6.52& -  & -  & -  & -  & -  &4.15& -  &4.91& -  &6.65& -  & -  & -  & -  & -  & -  \\
       &     &   &  1&0.04  & -  & -  & -  & -  &0.07& -  &0.12& -  & -  & -  & -  & -  &0.08& -  &0.13& -  &0.06& -  & -  & -  & -  & -  & -  \\
       &     &   &   &(3)   & -  & -  &(1) & -  &(2) & -  &(5) & -  & -  & -  & -  & -  &(3) & -  &(2) & -  &(16)& -  & -  & -  & -  & -  & -  \\ \hline
  46886&12900&3.8& 18&10.56 &7.84&7.42&8.54&7.87&6.52&6.33&6.63&5.73&6.67&5.14&6.47&4.00&5.53& -  &6.30&7.37&6.66&6.09&5.85& -  & -  & -  & -  \\
       &     &   &  1&0.06  &0.16&0.55& -  & -  &0.10& -  &0.17& -  &0.40& -  & -  &0.01&0.14& -  &0.11&0.13&0.11&0.13& -  & -  & -  & -  & -  \\
       &     &   &   &(4)   &(2) &(2) &(1) &(1) &(2) &(1) &(8) &(1) &(8) &(1) &(1) &(2) &(16)& -  &(19)&(23)&(23)&(2) &(1) & -  & -  & -  & -  \\ \hline
  47022&13000&3.8&123&10.98 & -  & -  &8.69& -  &7.58& -  &7.26& -  &7.46& -  & -  & -  &5.20& -  &5.65& -  &7.19& -  & -  & -  & -  & -  & -  \\
       &     &   &  8&0.07  & -  & -  & -  & -  & -  & -  &0.08& -  &0.17& -  & -  & -  & -  & -  &0.03& -  &0.18& -  & -  & -  & -  & -  & -  \\
       &     &   &   &(3)   & -  & -  &(1) & -  &(1) & -  &(2) & -  &(2) & -  & -  & -  &(1) & -  &(2) & -  &(4) & -  & -  & -  & -  & -  & -  \\ \hline
  47221&13800&4.0&189&11.10 & -  & -  & -  & -  &8.03& -  &7.33& -  &6.75& -  & -  & -  &5.51& -  & -  & -  &7.18& -  & -  & -  & -  & -  & -  \\
       &     &   & 13&0.01  & -  & -  & -  & -  &0.07& -  &0.15& -  &0.62& -  & -  & -  &0.14& -  & -  & -  &0.36& -  & -  & -  & -  & -  & -  \\
       &     &   &   &(2)   & -  & -  & -  & -  &(2) & -  &(2) & -  &(2) & -  & -  & -  &(2) & -  & -  & -  &(3) & -  & -  & -  & -  & -  & -  \\ \hline
  47257&10100&4.0& 48&10.80 & -  & -  &8.57& -  &7.52& -  &7.18& -  & -  & -  &5.83&2.99&4.53& -  &5.42&4.52&7.07& -  & -  & -  & -  & -  &2.13\\
       &     &   &  2&  -   & -  & -  & -  & -  &0.11& -  &0.08& -  & -  & -  &0.27& -  &0.10& -  &0.07& -  &0.09& -  & -  & -  & -  & -  & -  \\
       &     &   &   &(1)   & -  & -  &(1) & -  &(3) & -  &(4) & -  & -  & -  &(2) &(1) &(15)& -  &(6) &(1) &(21)& -  & -  & -  & -  & -  &(1) \\ \hline
  47272&10000&4.0& 21&10.74 &8.72& -  &8.55& -  &7.38&6.66&7.35& -  & -  & -  &6.17&2.77&4.69& -  &5.75& -  &7.36&6.42& -  &2.84&2.67& -  &2.98\\
       &     &   &  1&  -   & -  & -  &0.15& -  &0.12& -  &0.01& -  & -  & -  &0.03&0.19&0.09& -  &0.07& -  &0.10&0.28& -  &0.09&0.38& -  &0.04\\
       &     &   &   &(1)   &(1) & -  &(2) & -  &(6) &(1) &(4) & -  & -  & -  &(4) &(4) &(19)& -  &(8) & -  &(51)&(2) & -  &(2) &(2) & -  &(2) \\ \hline
  47278&11500&4.1& 38&10.73 & -  & -  &8.50& -  &7.18& -  &7.12& -  &7.00& -  &6.26& -  &5.51& -  &6.22&6.55&7.07& -  & -  & -  & -  & -  & -  \\
       &     &   &  2&0.06  & -  & -  & -  & -  &0.03& -  &0.10& -  & -  & -  & -  & -  &0.13& -  &0.09& -  &0.19& -  & -  & -  & -  & -  & -  \\
       &     &   &   &(4)   & -  & -  &(1) & -  &(2) & -  &(4) 7 -  &(1) & -  &(1) & -  &(15)& -  &(6) &(1) &(19)& -  & -  & -  & -  & -  & -  \\ \hline
  47431&12200&3.6&110&11.16 & -  & -  &8.87& -  &7.99& -  &7.35& -  & -  & -  & -  & -  &5.03& -  &5.53& -  &7.28& -  & -  & -  & -  & -  & -  \\
       &     &   &  9&0.01  & -  & -  &0.19& -  &0.17& -  &0.19& -  & -  & -  & -  & -  &0.12& -  & -  & -  &0.08& -  & -  & -  & -  & -  & -  \\
       &     &   &   &(2)   & -  & -  &(2) & -  &(2) & -  &(3) & -  & -  & -  & -  & -  &(2) & -  &(1) & -  &(8) & -  & -  & -  & -  & -  & -  \\ \hline
  47756&13300&3.6& 27&10.25 &8.67& -  &8.90& -  &7.49&5.40&8.21&5.42&6.90& -  & -  & -  &6.73& -  &7.00& -  &8.45&5.95& -  &4.30& -  & -  & -  \\
       &     &   &  1&0.12  & -  & -  & -  & -  &0.20& -  &0.04&0.13&0.08& -  & -  & -  &0.20& -  &0.12& -  &0.10& -  & -  & -  & -  & -  & -  \\
       &     &   &   &(2)   &(1) & -  &(1) & -  &(2) &(1) &(6) &(2) &(4) & -  & -  & -  &(8) & -  &(12)& -  &(46)&(1) & -  &(1) & -  & -  & -  \\ \hline
  47759&10200&3.9&100&11.18 & -  & -  &8.23& -  &7.58& -  & -  & -  & -  & -  & -  & -  &5.66& -  &7.59& -  &8.39& -  & -  & -  & -  & -  & -  \\
       &     &   &  4&  -   & -  & -  & -  & -  & -  & -  & -  & -  & -  & -  & -  & -  &0.06& -  &0.17& -  &0.08& -  & -  & -  & -  & -  & -  \\
       &     &   &   &(1)   & -  & -  &(1) & -  &(1) & -  & -  & -  & -  & -  & -  & -  &(2) & -  &(2) & -  &(7) & -  & -  & -  & -  & -  & -  \\ \hline
  47964&12300&3.2& 54&10.96 &8.16& -  &8.83&7.99&7.30& -  &7.01& -  &6.98& -  &6.59& -  &5.33& -  &5.84&5.71&7.09&6.22& -  & -  & -  & -  & -  \\
       &     &   &3  &0.03  & -  & -  & -  & -  &0.04& -  &0.04& -  &0.19& -  & -  & -  &0.20& -  &0.13&0.25&0.11&0.26& -  & -  & -  & -  & -  \\
       &     &   &   &(7)   &(1) & -  &(1) &(1) &(3) & -  &(5) & -  &(8) & -  &(1) & -  &(4) & -  &(4) &(2) &(18)&(2) & -  & -  & -  & -  & -  \\ \hline
  48212&12000&3.3&173&10.98 & -  & -  &8.88& -  &7.66& -  &7.30& -  &6.82& -  & -  & -  &5.26& -  &5.74& -  &7.24& -  & -  & -  & -  & -  & -  \\
       &     &   &  7&0.07  & -  & -  &0.05& -  & -  & -  &0.06& -  & -  & -  & -  & -  &0.26& -  & -  & -  &0.20& -  & -  & -  & -  & -  & -  \\
       &     &   &   &(2)   & -  & -  &(2) & -  &(1) & -  &(2) & -  &(1) & -  & -  & -  &(2) & -  &(1) & -  &(6) & -  & -  & -  & -  & -  & -  \\ \hline
  48497&14000&4.0& 13&11.01 &8.32&7.41&8.87&8.32&7.30&6.10&7.24& -  &6.97& -  &5.98& -  &4.88& -  &5.06& -  &7.06&5.85& -  &2.16& -  & -  & -  \\
       &     &   &  1&0.05  &0.09&0.49&0.11&0.05&0.20& -  &0.08& -  &0.10& -  &0.03& -  &0.03& -  &0.04& -  &0.14&0.07& -  & -  & -  & -  & -  \\
       &     &   &   &(5)   &(3) &(2) &(3) &(2) &(3) &(1) &(8) & -  &(11)& -  &(2) & -  &(2) & -  &(2) & -  &(33)&(2) & -  &(1) & -  & -  & -  \\ \hline
  48808&11900&3.2& 65&11.11 &7.77& -  &8.47& -  &7.14& -  &6.68& -  &6.99& -  & -  & -  &4.70& -  &4.95& -  &6.75&5.69& -  & -  & -  & -  & -  \\
       &     &   &  4&0.05  & -  & -  &0.08& -  &0.14& -  &0.07& -  &0.02& -  & -  & -  &0.11& -  &0.15& -  &0.19&0.09& -  & -  & -  & -  & -  \\
       &     &   &   &(7)   &(1) & -  &(2) & -  &(3) & -  &(7) & -  &(2) & -  & -  & -  &(5) & -  &(6) & -  &(17)&(2) & -  & -  & -  & -  & -  \\ \hline
  48957&12500&3.2& 34&10.91 &8.01& -  & -  & -  &7.17& -  &6.87& -  &6.35& -  & -  & -  &4.85& -  &5.07& -  &6.92& -  & -  & -  & -  & -  & -  \\
       &     &   &  4&0.09  & -  & -  & -  & -  & -  & -  &0.07& -  &0.44& -  & -  & -  &0.02& -  &0.06& -  &0.16& -  & -  & -  & -  & -  & -  \\
       &     &   &   &(4)   &(1) & -  & -  & -  &(1) & -  &(4) & -  &(2) & -  & -  & -  &(2) & -  &(2) & -  &(13)& -  & -  & -  & -  & -  & -  \\ \hline
  49123&10300&3.6& 47&10.97 &8.48&7.63&8.67& -  &7.37& -  &7.18& -  & -  & -  &6.14&3.10&4.45& -  &5.21& -  &6.84&5.88& -  & -  &2.87& -  &2.59\\
       &     &   &  2&0.04  &0.23& -  & -  & -  &0.04& -  &0.08& -  & -  & -  &0.05& -  &0.17& -  &0.18& -  &0.14& -  & -  & -  & -  & -  & -  \\
       &     &   &   &(2)   &(2) &(1) &(1) & -  &(2) & -  &(5) & -  & -  & -  &(2) &(1) &(9) & -  &(4) & -  &(16)&(1) & -  & -  &(1) & -  &(1) \\ \hline
  49481&12500&3.3& 11&10.73 &8.26&7.69&8.82&7.80&7.16&5.82&7.28&6.01&6.96& -  &5.92&3.32&4.91& -  &5.62&5.64&7.17&5.61& -  &2.45& -  & -  & -  \\
       &     &   &  1&  -   &0.17& -  &0.03&0.18&0.17&0.27&0.11&0.24&0.16& -  &0.12& -  &0.12& -  &0.10&0.37&0.11&0.19& -  & -  & -  & -  & -  \\
       &     &   &   &(1)   &(4) &(1) &(3) &(2) &(3) &(2) &(12)&(3) &(18)& -  &(2) &(1) &(16)& -  &(12)&(2) &(81)&(5) & -  &(1) & -  & -  & -  \\ \hline
  49643&14100&4.0&257&11.09 &8.35& -  & -  & -  &7.26& -  &7.80& -  &8.48& -  & -  & -  &5.17& -  & -  & -  &6.72& -  & -  & -  & -  & -  & -  \\
       &     &   & 22&0.06  & -  & -  & -  & -  &0.66& -  & -  & -  &0.66& -  & -  & -  &0.41& -  & -  & -  &0.04& -  & -  & -  & -  & -  & -  \\
       &     &   &   &(4)   &(1) & -  & -  & -  &(2) & -  &(1) & -  &(2) & -  & -  & -  &(2) & -  & -  & -  &(2) & -  & -  & -  & -  & -  & -  \\ \hline
  49711&11300&3.4&212&11.04 & -  & -  &8.81& -  &7.54& -  &7.53& -  &7.46& -  & -  & -  &4.39& -  &5.36& -  &7.29& -  & -  & -  & -  & -  & -  \\
       &     &   &  8&  -   & -  & -  & -  & -  &0.13& -  &0.08& -  & -  & -  & -  & -  &0.04& -  &0.02& -  &0.18& -  & -  & -  & -  & -  & -  \\
       &     &   &   &(1)   & -  & -  &(1) & -  &(2) & -  &(2) & -  &(1) & -  & -  & -  &(2) & -  &(2) & -  &(6) & -  & -  & -  & -  & -  & -  \\ \hline
  49713&12000&3.8& 50&10.66 & -  & -  &8.76& -  &6.55& -  &8.46& -  &7.61& -  &6.61&3.31&5.78& -  &7.85&6.74&8.50& -  & -  & -  & -  & -  & -  \\
       &     &   &  2&0.39  & -  & -  &0.03& -  & -  & -  &0.10& -  &0.38& -  &0.02& -  &0.20& -  &0.19&0.13&0.13& -  & -  & -  & -  & -  & -  \\
       &     &   &   &(2)   & -  & -  &(2) & -  &(1) & -  &(5) & -  &(2) & -  &(2) &(1) &(8) & -  &(26)&(2) &(27)& -  & -  & -  & -  & -  & -  \\ \hline
  49886&13000&4.0& 11&  -   &7.94& -  &8.74& -  &6.95&5.10&7.50&7.30&5.72& -  &6.66& -  &5.33& -  &5.94& -  &8.34&6.88& -  & -  & -  & -  & -  \\
       &     &   &  1&  -   & -  & -  &0.05& -  &0.19& -  &0.07&0.09&0.30& -  &0.06& -  &0.18& -  &0.10& -  &0.09&0.12& -  & -  & -  & -  & -  \\
       &     &   &   &      &(1) & -  &(3) & -  &(2) &(1) &(11)&(2) &(3) & -  &(2) & -  &(14)& -  &(12)& -  &(177)&(4)& -  & -  & -  & -  & -  \\ \hline
  49935&15200&4.0&100&10.59 &7.43& -  & -  & -  &7.59& -  &6.86& -  &6.74& -  &4.68& -  & -  & -  & -  & -  &7.49& -  & -  & -  & -  & -  & -  \\
       &     &   & 14&0.03  & -  & -  & -  & -  &0.03& -  &0.09& -  &0.17& -  & -  & -  & -  & -  & -  & -  &0.19& -  & -  & -  & -  & -  & -  \\
       &     &   &   &(5)   &(1) & -  & -  & -  &(2) & -  &(3) & -  &(5) & -  &(1) & -  & -  & -  & -  & -  &(8) & -  & -  & -  & -  & -  & -  \\ \hline
  50251&12000&3.2& 15&10.92 &8.32&7.81&8.73&8.02&7.54&6.06&7.30&5.60&6.91& -  &6.37& -  &4.55&4.22&5.57&5.99&7.20&6.09& -  &2.58&3.09& -  &2.24\\
       &     &   &  1&0.06  &0.09&0.06&0.12&0.20&0.19&0.05&0.09&0.08&0.17& -  &0.65& -  &0.13&0.08&0.11&0.07&0.13&0.11& -  & -  & -  & -  & -  \\
       &     &   &   &(4)   &(4) &(2) &(4) &(2) &(9) &(2) &(13)&(2) &(15)& -  &(2) & -  &(13)&(2) &(15)&(2) &(77)&(5) & -  &(1) &(1) & -  &(1) \\ \hline
  50252&10600&3.7&195&11.30 & -  & -  &8.14& -  &7.21& -  &6.54& -  & -  & -  & -  & -  &4.08& -  & -  & -  &6.61& -  & -  & -  & -  & -  & -  \\
       &     &   & 10&0.06  & -  & -  & -  & -  &0.79& -  &0.29& -  & -  & -  & -  & -  &0.72& -  & -  & -  &0.11& -  & -  & -  & -  & -  & -  \\
       &     &   &   &(2)   & -  & -  &(1) & -  &(2) & -  &(2) & -  & -  & -  & -  & -  &(2) & -  & -  & -  &(4) & -  & -  & -  & -  & -  & -  \\ \hline
  50513&11200&4.0&121&11.03 & -  & -  &8.28& -  &7.00& -  &6.71& -  & -  & -  & -  & -  &4.33& -  & -  & -  &6.71& -  & -  & -  & -  & -  & -  \\
       &     &   &  6&0.04  & -  & -  & -  & -  &0.08& -  &0.05& -  & -  & -  & -  & -  &0.67& -  & -  & -  &0.15& -  & -  & -  & -  & -  & -  \\
       &     &   &   &(2)   & -  & -  &(1) & -  &(2) & -  &(2) & -  & -  & -  & -  & -  &(2) & -  & -  & -  &(3) & -  & -  & -  & -  & -  & -  \\ \hline
  50751&13500&3.8&243&11.10 & -  & -  & -  & -  &7.61& -  &6.46& -  &6.46& -  & -  & -  & -  & -  & -  & -  &6.66& -  & -  & -  & -  & -  & -  \\
       &     &   & 14&0.07  & -  & -  & -  & -  &0.05& -  &0.13& -  &0.31& -  & -  & -  & -  & -  & -  & -  &0.04& -  & -  & -  & -  & -  & -  \\
       &     &   &   &(6)   & -  & -  & -  & -  &(2) & -  &(2) & -  &(2) & -  & -  & -  & -  & -  & -  & -  &(2) & -  & -  & -  & -  & -  & -  \\ \hline
  51079&11700&3.2&161&11.06 & -  & -  & -  & -  &7.74& -  &6.95& -  &6.35& -  & -  & -  &4.17& -  &6.54& -  &7.10& -  & -  & -  & -  & -  & -  \\
       &     &   &  8&0.02  & -  & -  & -  & -  & -  & -  &0.30& -  &0.39& -  & -  & -  &0.35& -  &0.16& -  &0.11& -  & -  & -  & -  & -  & -  \\
       &     &   &   &(3)   & -  & -  & -  & -  &(1) & -  &(2) & -  &(2) & -  & -  & -  &(2) & -  &(2) & -  &(3) & -  & -  & -  & -  & -  & -  \\ \hline
  52312&13000&3.4&175&10.74 & -  & -  &9.05& -  &6.55& -  &7.31& -  &7.31& -  &6.76& -  &5.47& -  &6.40& -  &7.40& -  & -  & -  & -  & -  & -  \\
       &     &   & 10&0.04  & -  & -  &0.20& -  &0.01& -  &0.05& -  & -  & -  & -  & -  &0.47& -  & -  & -  &0.13& -  & -  & -  & -  & -  & -  \\
       &     &   &   &(2)   & -  & -  &(2) & -  &(2) & -  &(2) & -  &(1) & -  &(1) & -  &(2) & -  &(1) & -  &(8) & -  & -  & -  & -  & -  & -  \\ \hline
  53004&11600&4.0& 58&10.78 &8.21& -  &8.82& -  &7.58& -  &7.13& -  &7.13& -  & -  &2.92&5.08& -  &5.94&7.62&7.03& -  & -  & -  & -  & -  & -  \\
       &     &   &  8&0.03  & -  & -  & -  & -  &0.03& -  &0.06& -  &0.11& -  & -  & -  &0.14& -  &0.05& -  &0.14& -  & -  & -  & -  & -  & -  \\
       &     &   &   &(2)   &(1) & -  &(1) & -  &(2) & -  &(4) & -  &(2) & -  & -  &(1) &(8) & -  &(4) &(1) &(11)& -  & -  & -  & -  & -  & -  \\ \hline
  53083&11600&3.5& 90&10.93 & -  & -  &8.86& -  &7.27& -  &6.89& -  &7.21& -  & -  & -  &4.55& -  &5.27& -  &6.97& -  & -  & -  & -  & -  & -  \\
       &     &   &  7&0.10  & -  & -  &0.03& -  &0.03& -  &0.15& -  &0.80& -  & -  & -  &0.03& -  &0.20& -  &0.15& -  & -  & -  & -  & -  & -  \\
       &     &   &   &(5)   & -  & -  &(2) & -  &(2) & -  &(3) & -  &(2) & -  & -  & -  &(3) & -  &(3) & -  &(10)& -  & -  & -  & -  & -  & -  \\ \hline
  53851&13600&3.7& 40&10.36 &7.74& -  & -  & -  &5.77& -  &7.71& -  &6.20& -  & -  & -  &5.16& -  &5.87& -  &7.55& -  & -  &3.61& -  & -  & -  \\
       &     &   &  3&0.09  & -  & -  & -  & -  & -  & -  &0.13& -  &0.19& -  & -  & -  &0.08& -  &0.15& -  &0.19& -  & -  &0.13& -  & -  & -  \\
       &     &   &   &(3)   &(1) & -  & -  & -  &(1) & -  &(8) & -  &(4) & -  & -  & -  &(2) & -  &(5) & -  &(31)& -  & -  &(2) & -  & -  & -  \\ \hline
  54929&10000&3.3& 81&11.07 & -  & -  &8.78& -  &7.24&6.41&7.27& -  & -  & -  &6.10& -  &4.47& -  &5.47&4.85&7.01& -  & -  & -  & -  & -  &2.23\\
       &     &   &  3&0.01  & -  & -  &0.01& -  &0.06& -  &0.03& -  & -  & -  & -  & -  &0.16& -  &0.06& -  &0.11& -  & -  & -  & -  & -  & -  \\
       &     &   &   &(2)   & -  & -  &(2) & -  &(4) &(1) &(3) & -  & -  & -  &(1) & -  &(12)& -  &(6) &(1) &(15)& -  & -  & -  & -  & -  &(1) \\ \hline
  55362&13000&4.0& 53&10.40 & -  & -  & -  & -  &5.84& -  &6.02& -  &5.18& -  & -  & -  &5.85& -  &5.68&6.24&6.99& -  & -  & -  & -  & -  & -  \\
       &     &   &  6&0.10  & -  & -  & -  & -  & -  & -  &0.01& -  &0.51& -  & -  & -  &0.38& -  &0.22&0.07&0.19& -  & -  & -  & -  & -  & -  \\
       &     &   &   &(2)   & -  & -  & -  & -  &(1) & -  &(2) & -  &(2) & -  & -  & -  &(2) & -  &(2) &(3) &(13)& -  & -  & -  & -  & -  & -  \\ \hline
  55793&11900&3.5&108&11.05 & -  & -  &8.75& -  &7.68& -  &7.06& -  &6.97& -  & -  & -  &4.34& -  &5.45& -  &7.04& -  & -  & -  & -  & -  & -  \\
       &     &   &  7&0.07  & -  & -  & -  & -  &0.14& -  &0.08& -  &0.06& -  & -  & -  &0.25& -  &0.12& -  &0.18& -  & -  & -  & -  & -  & -  \\
       &     &   &   &(5)   & -  & -  &(1) & -  &(3) & -  &(3) & -  &(2) & -  & -  & -  &(2) & -  &(2) & -  &(11)& -  & -  & -  & -  & -  & -  \\ \hline
  56006&14800&3.7&203&11.20 &8.34& -  & -  &8.06&7.60& -  &7.53& -  &7.08& -  &4.47& -  & -  & -  & -  & -  &7.96&7.00& -  & -  & -  & -  & -  \\
       &     &   & 10&0.07  &0.08& -  & -  & -  &0.08& -  &0.31& -  &0.03& -  & -  & -  & -  & -  & -  & -  &0.80& -  & -  & -  & -  & -  & -  \\
       &     &   &   &(7)   &(2) & -  & -  &(1) &(2) & -  &(2) & -  &(2) & -  &(1) & -  & -  & -  & -  & -  &(2) &(1) & -  & -  & -  & -  & -  \\ \hline
  56446&12800&3.5&195&11.03 &8.53& -  & -  &7.77&7.20& -  &7.03& -  &7.35& -  & -  & -  &5.54& -  & -  & -  &7.11& -  & -  & -  & -  & -  & -  \\
       &     &   & 15&0.05  & -  & -  & -  & -  &0.83& -  &0.01& -  &0.11& -  & -  & -  & -  & -  & -  & -  &0.16& -  & -  & -  & -  & -  & -  \\
       &     &   &   &(2)   & -  & -  & -  &(1) &(2) & -  &(2) & -  &(3) & -  & -  & -  &(1) & -  & -  & -  &(4) & -  & -  & -  & -  & -  & -  \\ \hline
  56610&13700&3.7&100&10.29 & -  & -  & -  & -  &6.84& -  &7.47& -  &7.17& -  & -  & -  &5.72& -  &6.38& -  &8.07& -  & -  & -  & -  & -  & -  \\
       &     &   & 10&0.03  & -  & -  & -  & -  & -  & -  &0.14& -  &0.09& -  & -  & -  &0.18& -  &0.16& -  &0.14& -  & -  & -  & -  & -  & -  \\
       &     &   &   &(3)   & -  & -  & -  & -  &(1) & -  &(6) & -  &(2) & -  & -  & -  &(4) & -  &(4) & -  &(13)& -  & -  & -  & -  & -  & -  \\ \hline
  56613&13300&4.0& 97&11.00 &8.04& -  & -  & -  &7.66& -  &7.28& -  &7.11& -  &5.98& -  &5.08& -  & -  & -  &7.27&6.22& -  & -  & -  & -  & -  \\
       &     &   &  8&0.07  & -  & -  & -  & -  &0.25& -  &0.15& -  &0.13& -  & -  & -  &0.74& -  & -  & -  &0.09& -  & -  & -  & -  & -  & -  \\
       &     &   &   &(3)   &(1) & -  & -  & -  &(2) & -  &(2) & -  &(4) & -  &(1) & -  &(2) & -  & -  & -  &(5) &(1) & -  & -  & -  & -  & -  \\ \hline
  57275& 9600&3.5&132&11.05 &7.73& -  &8.65& -  &6.99& -  &6.98& -  & -  & -  &6.14& -  &4.45& -  & -  & -  &6.97& -  & -  & -  & -  & -  & -  \\
       &     &   &  9&  -   & -  & -  & -  & -  &0.07& -  &0.25& -  & -  & -  & -  & -  &0.03& -  & -  & -  &0.18& -  & -  & -  & -  & -  & -  \\
       &     &   &   &(1)   &(1) & -  &(1) & -  &(2) & -  &(2) & -  & -  & -  &(1) & -  &(3) & -  & -  & -  &(7) & -  & -  & -  & -  & -  & -  \\ \hline
  57293&12000&4.0&160&11.04 & -  & -  &8.62& -  &8.16& -  &7.47& -  &8.02& -  & -  & -  &5.34& -  &6.38& -  &7.35& -  & -  & -  & -  & -  & -  \\
       &     &   & 17&0.32  & -  & -  & -  & -  &0.56& -  &0.23& -  & -  & -  & -  & -  &0.04& -  &0.12& -  &0.07& -  & -  & -  & -  & -  & -  \\
       &     &   &   &(2)   & -  & -  &(1) & -  &(2) & -  &(2) & -  &(1) & -  & -  & -  &(2) & -  &(2) & -  &(4) & -  & -  & -  & -  & -  & -  \\ \hline
 168202&10500&4.0& 71&11.01 & -  & -  &8.84& -  &7.55& -  &7.19& -  & -  & -  &6.66&3.01&5.01& -  &5.92&6.27&7.44& -  & -  & -  & -  & -  &3.39\\
       &     &   &  6&  -   & -  & -  &0.24& -  &0.01& -  &0.01& -  & -  & -  & -  & -  &0.10& -  &0.11& -  &0.17& -  & -  & -  & -  & -  & -  \\
       &     &   &   &(1)   & -  & -  &(2) & -  &(2) & -  &(2) & -  & -  & -  &(1) &(1) &(9) & -  &(6) &(1) &(17)& -  & -  & -  & -  & -  &(1) \\ \hline
 168932&10000&4.0& 38&10.26 &8.28&8.85&8.48& -  &7.93&6.75&7.52& -  & -  & -  &6.34&2.43&5.25&4.67&6.32&6.24&7.98& -  & -  &4.29&3.00&3.46&3.73\\
       &     &   &  1&0.12  &0.08& -  &0.04& -  &0.12& -  &0.20& -  & -  & -  &0.14&0.06&0.15& -  &0.07&0.10&0.08& -  & -  & -  &0.02& -  &0.10\\
       &     &   &   &(3)   &(2) &(1) &(2) & -  &(6) &(1) &(3) & -  & -  & -  &(2) &(2) &(20)&(1) &(11)&(4) &(42)& -  & -  &(1) &(2) &(1) &(3) \\ \hline
 169224&11200&4.0&152&11.01 & -  & -  &8.77& -  &7.68& -  &7.21& -  &7.15& -  & -  & -  &4.70& -  &5.55& -  &7.19& -  & -  & -  & -  & -  & -  \\
       &     &   & 11&  -   & -  & -  & -  & -  & -  & -  &0.12& -  & -  & -  & -  & -  &0.17& -  &0.07& -  &0.05& -  & -  & -  & -  & -  & -  \\
       &     &   &   &(1)   & -  & -  &(1) & -  &(1) & -  &(2) & -  &(1) & -  & -  & -  &(2) & -  &(2) & -  &(3) & -  & -  & -  & -  & -  & -  \\ \hline
 169225&10100&3.7&182&10.92 & -  & -  & -  & -  &7.86& -  &7.51& -  & -  & -  &7.11& -  &4.74& -  &5.42& -  &7.37& -  & -  & -  & -  & -  &2.83\\
       &     &   &  4&  -   & -  & -  & -  & -  & -  & -  & -  & -  & -  & -  & -  & -  &0.14& -  &0.38& -  &0.12& -  & -  & -  & -  & -  & -  \\
       &     &   &   &(1)   & -  & -  & -  & -  &(1) & -  &(1) & -  & -  & -  &(1) & -  &(3) & -  &(2) & -  &(3) & -  & -  & -  & -  & -  &(1) \\ \hline
 169512&12000&3.8&231&10.98 & -  & -  & -  & -  &7.99& -  &7.38& -  & -  & -  & -  & -  &4.79& -  &5.19& -  &7.38& -  & -  & -  & -  & -  & -  \\
       &     &   & 12&  -   & -  & -  & -  & -  &0.15& -  & -  & -  & -  & -  & -  & -  & -  & -  & -  & -  &0.14& -  & -  & -  & -  & -  & -  \\
       &     &   &   &(1)   & -  & -  & -  & -  &(2) & -  &(1) & -  & -  & -  & -  & -  &(1) & -  &(1) & -  &(2) & -  & -  & -  & -  & -  & -  \\ \hline
 169578&12000&3.7&259&11.09 & -  & -  & -  & -  &8.24& -  &7.52& -  & -  & -  & -  & -  &4.88& -  &5.28& -  &7.16& -  & -  & -  & -  & -  & -  \\
       &     &   & 14&0.10  & -  & -  & -  & -  &0.09& -  &0.23& -  & -  & -  & -  & -  &0.43& -  & -  & -  &0.04& -  & -  & -  & -  & -  & -  \\
       &     &   &   &(2)   & -  & -  & -  & -  &(2) & -  &(2) & -  & -  & -  & -  & -  &(2) & -  &(1) & -  &(3) & -  & -  & -  & -  & -  & -  \\ \hline
 170783&13800&3.0&295&11.11 & -  & -  & -  & -  &8.06& -  &7.62& -  &7.34& -  & -  & -  & -  & -  & -  & -  &6.99& -  & -  & -  & -  & -  & -  \\
       &     &   & 11&0.06  & -  & -  & -  & -  &0.46& -  & -  & -  &0.23& -  & -  & -  & -  & -  & -  & -  & -  & -  & -  & -  & -  & -  & -  \\
       &     &   &   &(2)   & -  & -  & -  & -  &(2) & -  &(1) & -  &(2) & -  & -  & -  & -  & -  & -  & -  &(1) & -  & -  & -  & -  & -  & -  \\ \hline
 170935&10300&3.0&271&10.79 & -  & -  & -  & -  &8.18& -  &7.36& -  & -  & -  & -  & -  &4.64& -  &5.66& -  &7.57& -  & -  & -  & -  & -  & -  \\
       &     &   & 16&  -   & -  & -  & -  & -  &0.09& -  & -  & -  & -  & -  & -  & -  & -  & -  & -  & -  &0.03& -  & -  & -  & -  & -  & -  \\
       &     &   &   &(1)   & -  & -  & -  & -  &(2) & -  &(1) & -  & -  & -  & -  & -  &(1) & -  &(1) & -  &(2) & -  & -  & -  & -  & -  & -  \\ \hline
 171247&12300&3.5& 68&  -   & -  & -  & -  & -  &7.48& -  &8.32& -  &6.75& -  & -  & -  &5.00& -  &6.05&6.85&7.72& -  & -  & -  & -  & -  & -  \\
       &     &   &  2&  -   & -  & -  & -  & -  &0.09& -  &0.12& -  & -  & -  & -  & -  &0.12& -  &0.03& -  &0.13& -  & -  & -  & -  & -  & -  \\
       &     &   &   &  -   & -  & -  & -  & -  &(4) & -  &(5) & -  &(1) & -  & -  & -  &(5) & -  &(6) &(1) &(18)& -  & -  & -  & -  & -  & -  \\ \hline
 171931&13000&3.7&283&10.93 & -  & -  & -  & -  &7.91& -  &6.82& -  &7.79& -  & -  & -  &5.71& -  &5.57& -  &6.94&7.67& -  & -  & -  & -  & -  \\
       &     &   & 30&0.02  & -  & -  & -  & -  &0.01& -  & -  & -  &0.65& -  & -  & -  &0.34& -  &0.32& -  & -  & -  & -  & -  & -  & -  & -  \\
       &     &   &   &(2)   & -  & -  & -  & -  &(2) & -  &(1) & -  &(2) & -  & -  & -  &(2) & -  &(2) & -  &(1) &(1) & -  & -  & -  & -  & -  \\ \hline
 172850&12000&3.9& 84&10.99 & -  & -  &8.66& -  &7.17& -  &6.81& -  & -  & -  &6.80& -  &4.70& -  &5.43& -  &6.89& -  & -  & -  & -  & -  & -  \\
       &     &   &  8&0.08  & -  & -  &0.07& -  &0.12& -  &0.07& -  & -  & -  & -  & -  &0.29& -  &0.02& -  &0.15& -  & -  & -  & -  & -  & -  \\
       &     &   &   &(5)   & -  & -  &(2) & -  &(4) & -  &(5) & -  & -  & -  &(1) & -  &(4) & -  &(2) & -  &(16)& -  & -  & -  & -  & -  & -  \\ \hline
 173673&13000&3.8& 29&  -   &7.92& -  & -  & -  &6.90&6.84&7.25&6.90&6.73& -  & -  & -  &5.73& -  &6.08& -  &8.06&6.20& -  & -  & -  & -  & -  \\
       &     &   &  2&  -   & -  & -  & -  & -  &0.16& -  &0.06&0.04&0.08& -  & -  & -  &0.18& -  &0.01& -  &0.13& -  & -  & -  & -  & -  & -  \\
       &     &   &   &  -   &(1) & -  & -  & -  &(2) &(1) &(6) &(2) &(3) & -  & -  & -  &(6) & -  &(3) & -  &(50)&(1) & -  & -  & -  & -  & -  \\ \hline
 174701&13500&3.9&178&11.13 & -  & -  &9.21& -  &7.83& -  &7.51& -  &7.22& -  & -  & -  &5.21& -  & -  & -  &7.69& -  & -  & -  & -  & -  & -  \\
       &     &   &  8&0.03  & -  & -  & -  & -  & -  & -  &0.04& -  & -  & -  & -  & -  &0.40& -  & -  & -  &0.11& -  & -  & -  & -  & -  & -  \\
       &     &   &   &(2)   & -  & -  & (1)& -  &(1) & -  &(2) & -  &(1) & -  & -  & -  &(2) & -  & -  & -  &(2) & -  & -  & -  & -  & -  & -  \\ \hline
 174836&10300&4.3&149&  -   & -  & -  &9.02& -  &7.87& -  &7.24& -  & -  & -  &6.88& -  &5.08& -  &5.65& -  &7.63& -  & -  & -  & -  & -  &3.49\\
       &     &   &  8&  -   & -  & -  & -  & -  & -  & -  &0.01& -  & -  & -  & -  & -  &0.16& -  &0.06& -  &0.07& -  & -  & -  & -  & -  & -  \\
       &     &   &   &  -   & -  & -  &(1) & -  &(1) & -  &(2) & -  & -  & -  &(1) & -  &(4) & -  &(3) & -  &(2) & -  & -  & -  & -  & -  &(1) \\ \hline
 174884&13200&3.6&113&11.04 &8.26& -  & -  & -  &7.81& -  &6.98& -  &7.02& -  & -  & -  &4.65& -  &5.66& -  &7.09& -  & -  & -  & -  & -  & -  \\
       &     &   &  4&0.10  & -  & -  & -  & -  &0.04& -  &0.52& -  &0.14& -  & -  & -  &0.13& -  &0.14& -  &0.12& -  & -  & -  & -  & -  & -  \\
       &     &   &   &(4)   &(1) & -  & -  & -  &(2) & -  &(2) & -  &(3) & -  & -  & -  &(2) & -  &(2) & -  &(6) & -  & -  & -  & -  & -  & -  \\ \hline
 176076&10400&3.7& 42&10.87 & -  & -  &8.79& -  &7.56& -  &7.32& -  & -  & -  &6.33&3.08&4.90& -  &5.72&4.68&7.28&6.61& -  &3.10& -  & -  &3.16\\
       &     &   &  2&0.11  & -  & -  &0.01& -  &0.15& -  &0.07& -  & -  & -  &0.21&0.21&0.11& -  &0.16& -  &0.12&0.02& -  & -  & -  & -  & -  \\
       &     &   &   &(2)   & -  & -  &(2) & -  &(4) & -  &(8) & -  & -  & -  &(2) &(4) &(14)& -  &(11)&(1) &(26)&(2) & -  &(1) & -  & -  &(1) \\ \hline
 176158&14500&3.7&126&11.12 &8.22& -  & -  & -  &7.06& -  &6.89& -  &6.61& -  &7.42& -  &5.78& -  &6.49& -  &6.41& -  & -  & -  & -  & -  & -  \\
       &     &   &  5&0.13  & -  & -  & -  & -  & -  & -  &0.08& -  & -  & -  & -  & -  & -  & -  & -  & -  &0.14& -  & -  & -  & -  & -  & -  \\
       &     &   &   &(4)   &(1) & -  & -  & -  &(1) & -  &(2) & -  &(1) & -  &(1) & -  &(1) & -  &(1) & -  &(2) & -  & -  & -  & -  & -  & -  \\ \hline
 176258&11000&3.9&303&11.00 & -  & -  & -  & -  &7.92& -  & -  & -  & -  & -  & -  & -  &5.03& -  & -  & -  &7.47& -  & -  & -  & -  & -  & -  \\
       &     &   &  8&0.13  & -  & -  & -  & -  &0.01& -  & -  & -  & -  & -  & -  & -  &0.07& -  & -  & -  & -  & -  & -  & -  & -  & -  & -  \\
       &     &   &   &(2)   & -  & -  & -  & -  &(2) & -  & -  & -  & -  & -  & -  & -  &(2) & -  & -  & -  &(1) & -  & -  & -  & -  & -  & -  \\ \hline
 177756&11700&4.0&177&10.91 & -  & -  & -  & -  &7.57& -  &7.64& -  & -  & -  & -  & -  &4.42& -  &5.82& -  &7.18& -  & -  & -  & -  & -  & -  \\
       &     &   &  6&0.09  & -  & -  & -  & -  & -  & -  & -  & -  & -  & -  & -  & -  &0.02& -  &0.60& -  &0.09& -  & -  & -  & -  & -  & -  \\
       &     &   &   &(2)   & -  & -  & -  & -  &(1) & -  &(1) & -  & -  & -  & -  & -  &(2) & -  &(2) & -  &(3) & -  & -  & -  & -  & -  & -  \\ \hline
 177880&14700&3.8& 56&11.03 &8.39& -  & -  &8.42&7.67& -  &7.03& -  &7.05& -  & -  & -  & -  & -  & -  & -  &7.14& -  & -  & -  & -  & -  & -  \\
       &     &   &  4&0.03  & -  & -  & -  &0.06&0.29& -  &0.12& -  &0.13& -  & -  & -  & -  & -  & -  & -  &0.10& -  & -  & -  & -  & -  & -  \\
       &     &   &   &(5)   &(1) & -  & -  &(2) &(2) & -  &(4) & -  &(6) & -  & -  & -  & -  & -  & -  & -  &(5) & -  & -  & -  & -  & -  & -  \\ \hline
 178744&14500&3.5&220&11.12 & -  & -  & -  & -  &7.92& -  &7.24& -  &7.47& -  & -  & -  &5.76& -  & -  & -  &6.51& -  & -  & -  & -  & -  & -  \\
       &     &   &  8&0.03  & -  & -  & -  & -  & -  & -  &0.75& -  & -  & -  & -  & -  & -  & -  & -  & -  & -  & -  & -  & -  & -  & -  & -  \\
       &     &   &   &(2)   & -  & -  & -  & -  &(1) & -  &(2) & -  &(1) & -  & -  & -  &(1) & -  & -  & -  &(1) & -  & -  & -  & -  & -  & -  \\ \hline
 179124&11600&3.2&298&11.23 & -  & -  & -  & -  &8.40& -  &7.55& -  & -  & -  & -  & -  &4.23& -  & -  & -  &6.22& -  & -  & -  & -  & -  & -  \\
       &     &   &  5&0.08  & -  & -  & -  & -  & -  & -  & -  & -  & -  & -  & -  & -  & -  & -  & -  & -  &0.10& -  & -  & -  & -  & -  & -  \\
       &     &   &   &(2)   & -  & -  & -  & -  &(1) & -  &(1) & -  & -  & -  & -  & -  &(1) & -  & -  & -  &(2) & -  & -  & -  & -  & -  & -  \\ \hline
 179761&12800&3.4& 17&11.15 &8.28&8.30&8.85&8.00&7.40& -  &7.42& -  &6.97& -  &6.41& -  &4.61& -  &5.54&6.26&7.30&6.07& -  &2.76& -  & -  & -  \\
       &     &   &  1&0.09  & -  & -  &0.15& -  &0.06& -  &0.13& -  &0.17& -  &0.22& -  &0.14& -  &0.07& -  &0.10&0.20& -  & -  & -  & -  & -  \\
       &     &   &   &(5)   &(1) &(1) &(2) &(1) &(5) & -  &(8) & -  &(10)& -  &(2) & -  &(7) & -  &(9) &(1) &(44)&(5) & -  &(1) & -  & -  & -  \\ \hline
 181440&11200&3.5& 58&10.90 & -  & -  & -  & -  &7.80& -  &7.24& -  & -  & -  & -  & -  &4.36& -  &5.09& -  &6.85& -  & -  & -  & -  & -  & -  \\
       &     &   &  1&0.12  & -  & -  & -  & -  &0.26& -  &0.59& -  & -  & -  & -  & -  &0.45& -  & -  & -  &0.08& -  & -  & -  & -  & -  & -  \\
       &     &   &   &(2)   & -  & -  & -  & -  &(2) & -  &(2) & -  & -  & -  & -  & -  &(2) & -  &(1) & -  &(5) & -  & -  & -  & -  & -  & -  \\ \hline
 181761&14600&3.6& 35&10.91 &7.61& -  &8.49&7.68&7.17& -  &6.45& -  &6.99& -  & -  & -  &5.45& -  &5.23& -  &7.12& -  & -  & -  & -  & -  & -  \\
       &     &   &  3&0.11  & -  & -  & -  & -  &0.30& -  &0.19& -  &0.01& -  & -  & -  &0.04& -  & -  & -  &0.17& -  & -  & -  & -  & -  & -  \\
       &     &   &   &(4)   &(1) & -  &(1) &(1) &(2) & -  &(3) & -  &(2) & -  & -  & -  &(2) & -  &(1) & -  &(5) & -  & -  & -  & -  & -  & -  \\ \hline
 182198&11450&3.5& 25&10.93 &8.38& -  &7.61& -  &7.48&6.56&7.34&5.28&7.06& -  &6.15& -  &4.61&4.55&5.65& -  &7.27&6.45& -  & -  & -  & -  &2.56\\
       &     &   &  1&0.03  &0.14& -  &0.24& -  &0.13&0.42&0.09&0.02&0.17& -  &0.12& -  &0.14&0.01&0.12& -  &0.16&0.12& -  & -  & -  & -  &0.01\\
       &     &   &   &(5)   &(8) & -  &(2) & -  &(22)&(6) &(14)&(2) &(20)& -  &(5) & -  &(18)&(2) &(30)& -  &(93)&(15)& -  & -  & -  & -  &(2) \\ \hline
\hline
\end{longtable}
\end{scriptsize}
\end{landscape}
}
\begin{acknowledgements}
The authors wish to thank the referee, Douglas Gies, for useful comments. This study was funded by means of a 6-month Visiting Postdoctoral Fellowship assigned to EN by the Fund for Scientific Research of Flanders (FWO) in the framework of FWO project G.0332.06. The authors are supported by the Research Council of Leuven University under grant GOA/2008/04 and by the European Helio- and Asteroseismology Network (HELAS), a major international collaboration funded by the European Commission's Sixth Framework Programme.
\end{acknowledgements}

{}

\begin{appendix} 
\section{Reliability of the method and effect of the uncertainties in the atmospheric parameters on the abundance results}
To assess the robustness of the procedures used to determine the chemical abundances, we computed a grid of synthetic spectra and compared the input values with the results derived by our method. The choice of parameters for the grid was dictated by the range in
effective temperatures of our sample. Typical surface gravity and microturbulence were adopted: \logg\,=\,4.0\,dex and \turb\,=\,2\,\kms.
We broadened the spectra with five different rotational velocities sampling the range covered by our stars (10--200\,\kms) to check whether our method recovers the true abundances even for rapid rotators. The parameters adopted for the grid of theoretical spectra are presented in Table\,\ref{table-ap}. Solar metallicity was assumed, but the individual abundances of some elements were slightly changed.

All the synthetic spectra were degraded by white, Gaussian noise to achieve a signal-to-noise ratio of about 100 typical of our data (the synthetic spectra also have a resolving power of 40\,000 matching that of the observed spectra).

\begin{table}
\caption{Stellar parameters and rotational velocites of the test synthetic spectra.}
\label{table-ap}
\centering
\begin{tabular}{c|c|c|c} 
\hline 
\teff~  & \logg~  & \turb~ & \vsini~ \\
 ~[K] &       & [\kms ]  & [\kms ] \\
\hline
10000 & 4.0 & 2 & 10, 50, 100, 150, 200\\
12000 & 4.0 & 2 & 10, 50, 100, 150, 200\\
14000 & 4.0 & 2 & 10, 50, 100, 150, 200\\
\hline
\end{tabular}
\end{table}

In Fig.\ref{app1}, we compare the obtained chemical abundances ($\log\varepsilon({\rm El})_{\rm obtained}$) with the input values adopted for the synthetic spectra ($\log\varepsilon({\rm El})_{\rm model}$). The differences are plotted with different symbols depending on the assumed rotational velocity. For clarity, the results for a given temperature are presented in separate panels. The discrepancies are in most cases lower than 0.1\,dex. For \teff\,=\,10\,000\,K, the largest differences are found for \vsini\,=\,200\,\kms, while for \teff\,=\,12\,000\,K they are lower than 0.1\, in all cases. For \teff\,=\,14\,000\,K, the largest discrepancies are found for \vsini\,=\,150\,\kms\, in the case of Ca and Ni. The lines are very weak in this case, and for \vsini\,=\,200\,\kms, they are no longer measurable.

   \begin{figure}
   \centering
   \includegraphics[width=9cm]{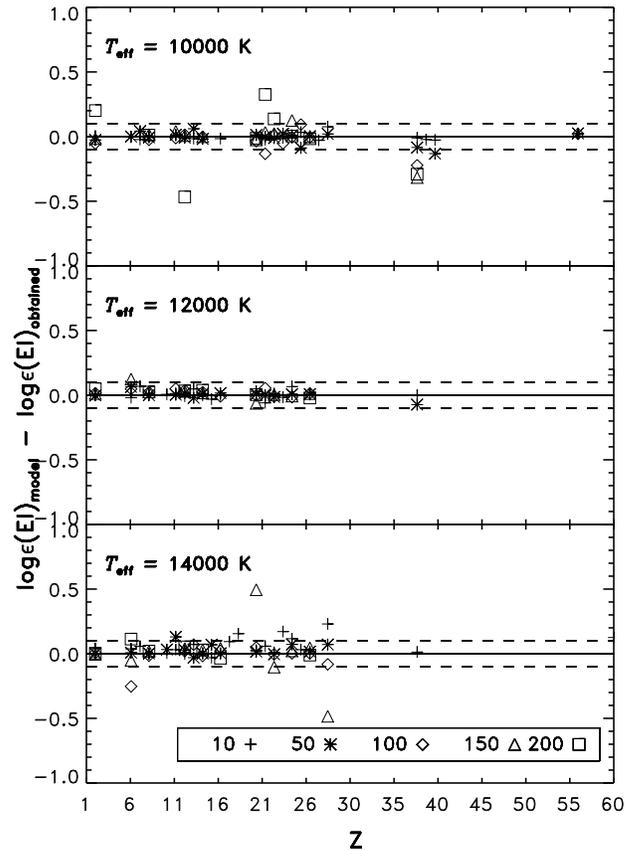}
      \caption{Differences between adopted and determined abundances for all assumed rotation velocities.}
   \label{app1}
   \end{figure}

   \begin{figure}
   \centering
   \includegraphics[width=9cm]{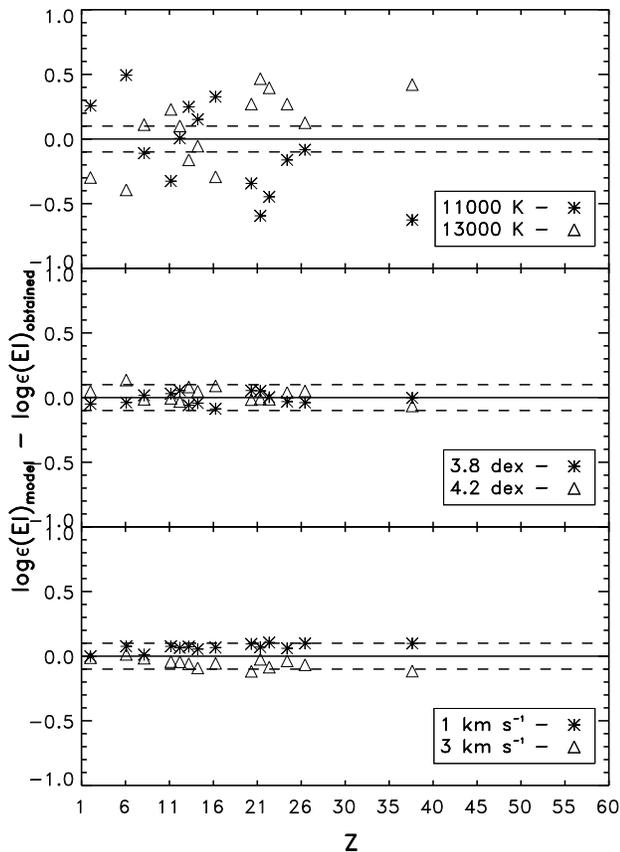}
      \caption{Differences between adopted and determined abundances assuming atmospheric models with an incorrect effective temperature (top panel), surface gravity (middle panel) or microturbulence (bottom panel).}
   \label{app2}
   \end{figure}

This test demonstrates the usefulness of our method for the determination of the abundance pattern of late B-type stars even at high rotation velocities. Only in the case of weak, blended features are the results prone to large uncertainties.

To estimate the extent to which the uncertainties in the atmospheric parameters influence the determined abundances, we considered a synthetic spectrum calculated for \teff\,=\,12\,000\,K, \logg\,=\,4.0, \turb\,=\,2\,\kms, and \vsini\,=\,50\,\kms. The abundances were derived based on this spectrum and by assuming atmospheric models that deviate by $\Delta$\teff\,=\,$\pm$1000 K, $\Delta$\logg\,=\,$\pm$0.2\,dex, and $\Delta$\turb\,=\,$\pm$1\,\kms\, from the reference values (the other two parameters were kept frozen when one was allowed to vary). Comparison of the results obtained with the reference atmosphere model allows us to assess the effect of typical errors in the effective temperature, surface gravity and microturbulence on the determined abundances.

We present in Fig.\ref{app2} the differences between the input and calculated abundances, assuming that the adopted atmospheric parameters (\teff, \logg, \turb) chosen are incorrect. An error in the effective temperature of 1000\,K has the largest effect on the determined chemical abundances. In this case, the discrepancies can exceed 0.5\,dex because the abundances are based on trace ions that are very sensitive to changes in \teff. The proper determination of \teff\, is crucial to obtain the correct abundance pattern of the star. For $\Delta \log g$\,=\,$\pm$ 0.2 and $\Delta$\turb\,=\,$\pm$ 1\,\kms, the effect is much smaller and the differences are typically below 0.1\,dex.

\end{appendix}

\end{document}